\documentclass[journal,a4paper]{IEEEtran}
\usepackage{cite}

\ifCLASSINFOpdf
\else
\fi
\ifCLASSOPTIONcompsoc
  \usepackage[caption=false,font=normalsize,labelfont=sf,textfont=sf]{subfig}
\else
  \usepackage[caption=false,font=footnotesize]{subfig}
\fi
\usepackage{stfloats}
\newcommand\blfootnote[1]{%
  \begingroup
  \renewcommand\thefootnote{}\footnote{#1}%
  \addtocounter{footnote}{-1}%
  \endgroup
}

\usepackage{tikz,pgfplots,pgfplotstable}
\pgfplotsset{scaled x ticks=false}
\pgfplotsset{
        compat=newest
    }
\usetikzlibrary{shadows,arrows,shapes}
\graphicspath{{./figures/}}

\usepackage[cmex10]{amsmath}
\usepackage{amssymb}
\usepackage{balance}
\DeclareMathOperator*{\argmax}{\mathrm{argmax}}

\usepackage[ddmmyyyy]{datetime}
\newcommand{\header}{PREPRINT, \today, \currenttime}

\begin{document}

\renewcommand{\arraystretch}{1.4} 
\title{Information Rates of Next-Generation Long-Haul Optical Fiber Systems Using Coded Modulation}
%
%
%

\author{Gabriele Liga,~\IEEEmembership{Student Member,~IEEE,}
        Alex~Alvarado,~\IEEEmembership{Senior Member,~IEEE,}
        Erik~Agrell,~\IEEEmembership{Senior Member,~IEEE,}
        and~Polina~Bayvel,~\IEEEmembership{Fellow,~IEEE},~\IEEEmembership{Senior Member,~OSA}
\thanks{}
\thanks{}
\thanks{}}

%
%

\markboth{\header}
{\header}
%



\maketitle
\begin{abstract}
A comprehensive study of the coded performance of long-haul spectrally-efficient WDM optical fiber transmission systems with different coded modulation decoding structures is presented. 
Achievable information rates are derived for three different square QAM formats and the optimal format is identified as a function of distance and specific decoder implementation. The four cases analyzed combine hard-decision (HD) or soft-decision (SD) decoding together with either a bit-wise or a symbol-wise demapper, the last two suitable for binary and nonbinary codes, respectively. The information rates achievable for each scheme are calculated based on the mismatched decoder principle. These quantities represent true indicators of the coded performance of the system for specific decoder implementations and when the modulation format and its input distribution are fixed. In combination with the structure of the decoder, two different receiver-side equalization strategies are also analyzed: electronic dispersion compensation and digital backpropagation. We show that, somewhat unexpectedly, schemes based on nonbinary HD codes can achieve information rates comparable to SD decoders and that, when SD is used, switching from a symbol-wise to a bit-wise decoder results in a negligible penalty. Conversely, from an information-theoretic standpoint, HD binary decoders are  shown to be unsuitable for spectrally-efficient, long-haul systems.        
\end{abstract}

\IEEEpeerreviewmaketitle

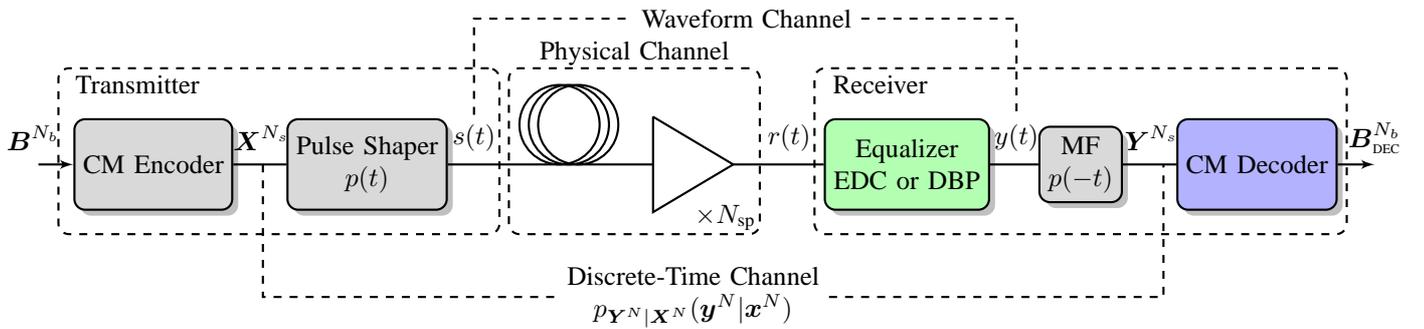
\begin{figure*}[!t]
\begin{tikzpicture}[>=latex',thick]
\tikzstyle{block} = [draw,fill=red!30,minimum size=3em,rounded corners];
\tikzstyle{ring} = [circle,draw,minimum size=3em];
\node[block,fill opacity=0,dashed,minimum width=5.8cm, minimum height=2.2cm] at (-35pt,5pt) (TX) {};
\node at (-88pt,30pt) {Transmitter};
\node[block,fill=gray!30,drop shadow,minimum width=1.8cm,minimum height=1.2cm] at (-82pt,0pt)(Enc){CM Encoder};
\node[block,fill=gray!30,drop shadow,align=center,minimum width=1.8cm,minimum height=1.2cm] at (-2pt,0pt)(PulseShape){Pulse Shaper \\ $p(t)$};
\node[ring] at (70pt,15pt) (Circ){};
\node[ring] at (73.5pt,15pt) (Circ){};
\node[ring] at (77pt,15pt) (Circ){};
\node[block,fill opacity=0,dashed,minimum width=3.3cm, minimum height=2.2cm] at (98pt,5pt) (Ch) {};
\node[block,fill opacity=0,dashed,minimum width=7cm, minimum height=2.2cm] at (265pt,5pt) (RX) {};
\draw (PulseShape) -- (105pt,0pt);

\node[align=center] at (98pt,42pt) (PHY) {Physical Channel};
\node[align=center] at (132pt,-20pt) (Nspans) {$\times N_{\text{sp}}$}; 
\node[align=center] at (-127pt,10pt) (Bits) {$\boldsymbol{B}^{N_b}$};
\node[align=center] at (-41.5pt,10pt) (TXSyms) {$\boldsymbol{X}^{N_s}$};
\node[align=center] at (38pt,10pt) (TXsig) {$s(t)$};
\node[align=center] at (156pt,10pt) (RXsig) {$r(t)$};
\node[align=center] at (241pt,10pt) (RXsig) {$y(t)$};
\node[align=center] at (291.5pt,10pt) (RXSyms) {$\boldsymbol{Y}^{N_s}$};
\node[align=center] at (375pt,10pt) (Bits) {$\boldsymbol{B}^{N_b}_{\text{\tiny DEC}}$};

\draw (105pt,18pt) -- (105pt,-18pt) -- (135pt,0pt) -- (105pt,18pt);
\node[block,fill=green!30,minimum height=1.2cm, drop shadow,align=center] at (200pt,0pt) (Eq) {Equalizer \\ EDC or DBP};
\draw (135pt,0pt) -- (Eq);
\node[block,minimum size=1cm,fill=gray!30,drop shadow,align=center] at (265pt,0pt) (MF) {MF \\ $p(-t)$};
\node[block,minimum width=1.3cm,minimum height=1.2cm,fill=blue!30,drop shadow] at (331pt,0pt) (Dec) {CM Decoder};

\node at (190pt,30pt) {Receiver};
\node[align=center] at (140pt,55pt) (WaveformCh) {Waveform Channel};
\node[align=center] at (120pt,-50pt) (ChLaw) {Discrete-Time Channel\\ $p_{\boldsymbol{Y}^N|\boldsymbol{X}^N}(\boldsymbol{y}^N|\boldsymbol{x}^N)$};
\draw [thick,->] (-125pt,0pt) -- (Enc);
\draw [thick] (Enc) -- (PulseShape);
\draw [thick] (Eq) -- (MF); 
\draw [thick] (MF) -- (Dec); 
\draw [dashed] (TXsig) -- (38pt,55pt) -- (WaveformCh) -- (241pt,55pt) -- (RXsig); 
\draw [dashed] (-41pt,0pt) -- (-41pt,-50pt) -- (ChLaw) -- (296pt,-50pt) -- (296pt,0pt); 
\draw [thick,->] (Dec) -- (375pt,0pt);

\end{tikzpicture}
\caption{General schematic of the optical communication system analyzed in this work.}
\label{fig:SystemSchematic}
\end{figure*} 
\section{Introduction}
\blfootnote{Research supported by the Engineering and Physical Sciences Research Council (EPSRC) through the programme grant UNLOC (EP/J017582/1).
Parts of this work were submitted for possible publication at the 2016 IEEE Photonics Conference.}
\blfootnote{G. Liga, A. Alvarado, and P. Bayvel are with the Optical Networks Group, Department of Electronic and Electrical Engineering, University College London, London WC1E 7JE, United Kingdom (email: gabriele.liga.11@ucl.ac.uk).}
\blfootnote{E. Agrell is with the Department of Signals and Systems, Chalmers University of Technology, SE-41296, Gothenburg, Sweden.}

~\IEEEPARstart{T}{he} demand for ever higher transmission rates in optical fiber transmission systems has led researchers to study the performance of transceivers based on sophisticated forward error correction (FEC) techniques. Next-generation long-haul transceivers will use powerful FEC and high-spectral-efficiency (SE) modulation formats, a combination known as coded modulation (CM). In order to provide reliable transmission, a FEC encoder maps blocks of information bits into longer blocks of coded bits that are sent through the channel at a nominal transmission rate. As a result, the information rate is, in general, lower than the nominal one by an amount that depends on the redundancy added by the FEC encoder, which in turn needs to be adjusted based on the quality of the channel. A key performance parameter of such systems is then the the maximum rate at which an optical communication system can be operated whilst maintaining reliable transmission of information. 

To have an estimate of this rate, a widely used approach in the optical communication literature is based on identifying a pre-FEC BER threshold, for which a specific high-performance FEC code can guarantee an error-free performance after decoding. The code rate of such a coding scheme, multiplied by the raw transmission data rate, is used to identify an achievable information rate (AIR) for that specific system configuration. On the other hand, information theory, founded by Shannon in his seminal paper \cite{Shannon1948}, shows that quantities such as the mutual information (MI) can precisely indicate what is the maximum information rate at which a code can ensure an arbitrarily small error probability \cite{Gallager1968,Cover2006}. Moreover, several recent works have showed that both the MI and the generalized mutual information (GMI) \cite{Caire1998,Szczecinski2015bit} are more reliable indicators than the pre-FEC BER of the performance of coded optical fiber systems, regardless of the specific channel used for transmission \cite{Leven2011,Schmalen2012,Schmalen2016,Alvarado2015III,Alvarado2015,Alvarado16a,MERL2016}.

The \emph{channel MI} (i.e., the MI including the channel memory) represents an upper limit on the AIRs for a given channel when a given modulation format is used and an optimum maximum likelihood (ML) decoder is used at the receiver. However, the implementation of such a decoder is prohibitive, both for complexity reasons and due to the lack of knowledge of the channel law.  Instead of the optimum decoder, more pragmatic CM decoders are usually employed. Typical CM decoder implementations used in optical communications neglect the channel memory \cite{Alvarado2015III} and are, thus, suboptimal. Furthermore, their design involves two degrees of freedom. Each degree of freedom presents two options: hard-decision (HD) vs. soft-decision (SD) decoding and bit-wise (BW) vs. symbol-wise (SW) demapping, effectively producing four different design options. The channel MI is not in general an AIR for any of these four suboptimal schemes. Indeed, the adopted decoding strategy has a major impact on the AIRs, which can potentially be significantly lower that the channel MI. 
A common approach to calculate AIRs for specific decoder implementations is based on two steps: i) the memory of the optical fiber channel is neglected and the MI is calculated for an equivalent memoryless channel; ii) the \textit{mismatched decoder} principle is used \cite{Merhav1994,Arnold2006,Colavolpe2011,Secondini2013}. Each of these two methods results in a lower bound on the channel MI.
In \cite{Essiambre2010} the memoryless MI was studied for coherent optical fibre systems using ring constellations. In \cite{Leven2011,Schmalen2012}, the same quantity was used in an experimental scenario as a system performance metric for an SD coded system. In \cite{Alvarado2015III} and \cite[Fig.~6]{Alvarado2015}, it was shown that when BW decoders are used, the GMI is a better metric to predict AIRs than the MI. The GMI has also been used to evaluate the performance of experimental optical systems in \cite{Maher2016,ALU2015,Eriksson2015}. The memoryless MI and the GMI were also shown to be good post-FEC BER predictors for SD-SW (nonbinary) and SD-BW decoders, in \cite{Schmalen2016} and \cite{Alvarado2015} respectively. Finally, a study comparing SD-SW and HD-BW AIRs for polarization multiplexed (PM) quadrature-amplitude modulation (QAM) formats (PM-16QAM and PM-64QAM) was presented in \cite{Fehenberger2015}, where electronic dispersion compensation (EDC) or digital backpropagation (DBP) are used at the receiver for a given transmission distance.

In this work, we extend the results in \cite{Fehenberger2015} adding, for the first time, AIRs for HD-SW decoders to the picture. Furthermore, we present a comprehensive comparison of the AIRs of the optical fiber channel for different CM decoder implementations and for all transmission distances of interest for mid-range/long-haul terrestrial and transoceanic optical fiber links. The AIRs are also compared for different equalization techniques and over different PM-$M$QAM formats with nominal SE above 4 bits/sym per polarization such as PM-16QAM, PM-64QAM, and PM-256QAM. The results in this paper show the design trade-offs in coded optical fiber systems where, for a given distance requirement,there is a trade off between transmission rates and transceiver complexity (modulation format, equalization, and decoding). To the best of our knowledge, this is the first time such an extensive study is performed for optical fiber communication systems.  
The paper is structured as follows: in Section~\ref{sec2}, the investigated system is first modeled and the different decoding strategies analyzed in this work are described; Section~\ref{sec3} discusses in a semi-tutorial style the information-theoretic quantities used to evaluate their performance and, as a reference, results are shown for the additive white Gaussian noise (AWGN) channel. In Section~\ref{sec4}, the numerical setup is explained and AIR results for the optical fiber channel are shown; finally in Section~\ref{sec5}, conclusions are drawn.
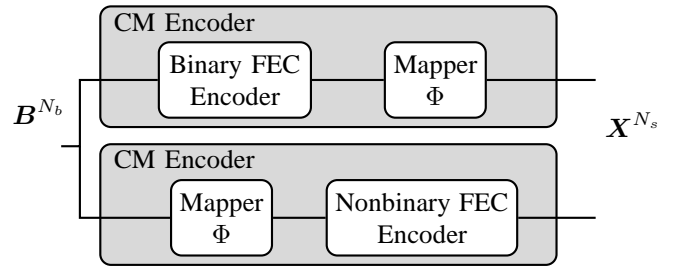
\begin{figure}[!t]
\begin{tikzpicture}
\tikzstyle{block} = [draw,minimum width=6cm,minimum height=1.6cm,rounded corners,thick,fill=white];
\node at (0pt,0pt) (Ref){};
\node[block,fill=gray!30] at (125pt,0pt) (CM_Enc1){};
\node[block,fill=gray!30] at (125pt,-52pt) (CM_Enc2){};
\node at (71pt,17pt) (EncLabel){CM Encoder};
\node at (71pt,-35pt) (EncLabel){CM Encoder};
\node[block,minimum width=2cm,minimum height=1cm,align=center] at (90pt,-5pt) (BinFEC){Binary FEC \\ Encoder};
\node[block,minimum width=1cm,minimum height=1cm,align=center] at (165pt,-5pt) (Map1){Mapper \\ $\Phi$};
\node[block,minimum width=1cm,minimum height=1cm,align=center] at (85pt,-57pt) (Map2){Mapper \\ $\Phi$};
\node[block,minimum width=2cm,minimum height=1cm,align=center] at (160pt,-57pt) (NBFEC){Nonbinary FEC \\ Encoder};
\draw[thick] (BinFEC) -- (Map1);
\draw[thick] (Map2) -- (NBFEC);
\node[anchor=south east] at (30pt,-25pt) (input){$\boldsymbol{B}^{N_b}$};
\node[anchor=south west] at (225pt,-30pt) (output){$\boldsymbol{X}^{N_s}$};
\draw[thick] (25pt,-30pt) -- (32pt,-30pt) -- (32pt,-5pt) -- (BinFEC);
\draw[thick] (32pt,-30pt) -- (32pt,-57pt) -- (Map2);
\draw[thick] (Map1) -- (225pt,-5pt);
\draw[thick] (NBFEC) -- (225pt,-57pt);
\end{tikzpicture}
\caption{Two different implementation alternatives for the CM encoder in Fig.~\ref{fig:SystemSchematic}.}
\label{fig:CM Encoder}
\end{figure}   

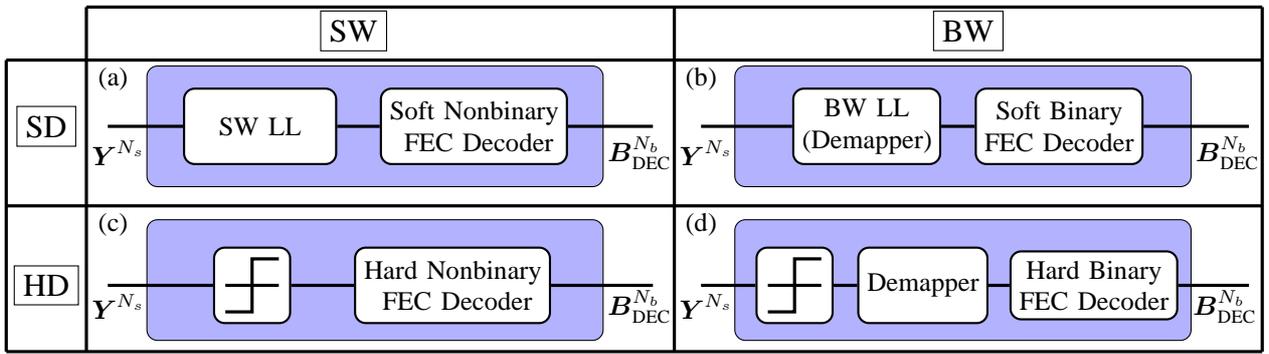
\begin{figure*}[!t]
\centering
\begin{tikzpicture}
\tikzstyle{rx} = [draw,minimum height=1.6cm,minimum width=6cm,fill=blue!30,rounded corners];
\tikzstyle{block} = [draw,minimum size=1cm,rounded corners,thick,fill=white];
\node at (-130pt,0pt) (ref){};

\node[rx] at (-2pt,30pt) (RX1){};
\node[block,align=center,minimum width=2cm,minimum height=1cm] at (-45pt,30pt) (SW-LLR){SW LL};
\node[block,align=center,minimum width=2cm] at (35pt,30pt) (NB-DEC2){Soft Nonbinary \\\ FEC Decoder};
\draw[black,very thick] (SW-LLR) -- (NB-DEC2);

\node[rx] at (218pt,30pt) (RX2){};
\node[block,align=center] at (182pt,30pt) (BW-LLR){BW LL \\ (Demapper)};
\node[block,align=center,minimum size=1cm] at (254pt,30pt) (BinFECDec){Soft Binary \\ FEC Decoder};
\draw[black,very thick] (BW-LLR) -- (BinFECDec);

\node[rx] at (-2pt,-28pt) (RX3){};
\node[block,align=center,minimum size=1cm] at (-48pt,-30pt) (threshold2){};
\draw[black,very thick] (-58pt,-40pt) -- (-48pt,-40pt) --(-48pt,-20pt) --(-38pt,-20pt);
\draw[black,very thick] (-58pt,-30pt) -- (-38pt,-30pt);
\node[block,align=center,minimum width=1.8cm] at (27pt,-30pt) (NB-DEC1){Hard Nonbinary \\\ FEC Decoder};
\draw[black,very thick] (threshold2) -- (NB-DEC1);

\node[rx] at (218pt,-28pt) (RX4){};
\node[block,align=center,minimum size=1cm] at (155pt,-30pt) (threshold1){};
\draw[black,very thick] (165pt,-20pt) -- (155pt,-20pt) --(155pt,-40pt) --(145pt,-40pt);
\draw[black,very thick] (145pt,-30pt) -- (165pt,-30pt);
\node[block,align=center,minimum size=1cm] at (203pt,-30pt) (BW-Dem){Demapper};
\node[block,align=center,minimum size=0.6cm] at (267pt,-30pt) (HD-Bin-Dec){Hard Binary \\ FEC Decoder};
\draw[black,very thick] (threshold1) -- (BW-Dem) -- (HD-Bin-Dec);

\draw[black,very thick] (-140pt,0pt) -- (330pt,0pt);
\draw[black,very thick] (110pt,75pt) -- (110pt,-55pt);
\draw[black,very thick] (-140pt,55pt) -- (-140pt,-55pt);
\draw[black,very thick] (-140pt,55pt) -- (330pt,55pt);
\draw[black,very thick] (-110pt,75pt) -- (-110pt,-55pt);
\draw[black,very thick] (-110pt,75pt) -- (330pt,75pt);
\draw[black,very thick]  (330pt,75pt) -- (330pt,-55pt);
\draw[black,very thick]  (-140pt,-55pt) -- (330pt,-55pt);
\node[draw] at (-125pt,30pt) {\large SD};
\node[draw] at (-125pt,-30pt) {\large HD};
\node[draw] at (-10pt,66pt) {\large SW};
\node[draw] at (220pt,66pt) {\large BW};
\node[] at (-98pt,20pt) {$\boldsymbol{Y}^{N_s}$};
\node[] at (97pt,20pt) {$\boldsymbol{B}^{N_b}_{\text{DEC}}$};
\node[] at (122pt,20pt) {$\boldsymbol{Y}^{N_s}$};
\node[] at (317pt,20pt) {$\boldsymbol{B}^{N_b}_{\text{DEC}}$};
\node[] at (-98pt,-38pt) {$\boldsymbol{Y}^{N_s}$};
\node[] at (97pt,-38pt) {$\boldsymbol{B}^{N_b}_{\text{DEC}}$};
\node[] at (122pt,-38pt) {$\boldsymbol{Y}^{N_s}$};
\node[] at (316pt,-38pt) {$\boldsymbol{B}^{N_b}_{\text{DEC}}$};
\draw[black,very thick] (-102pt,30pt)--(SW-LLR);
\draw[black,very thick] (NB-DEC2)--(102pt,30pt);
\draw[black,very thick] (120pt,30pt)--(BW-LLR);
\draw[black,very thick] (BinFECDec)--(325pt,30pt);
\draw[black,very thick](-102pt,-30pt)--(threshold2);
\draw[black,very thick](NB-DEC1)--(102pt,-30pt);
\draw[black,very thick](120pt,-30pt)--(threshold1);
\draw[black,very thick](HD-Bin-Dec)--(325pt,-30pt);

\node[] at (-100pt,48pt) {(a)};
\node[] at (120pt,48pt) {(b)};
\node[] at (-100pt,-7pt) {(c)};
\node[] at (120pt,-7pt) {(d)};
\end{tikzpicture}
\caption{The four CM decoder implementations analyzed in this work.}
\label{fig:Decoders}
\end{figure*}

\section{System Model}\label{sec2}
We consider the schematic diagram in Fig.~\ref{fig:SystemSchematic}, representing a generic multispan optical fiber communication system. Although in this work PM (4D) modulation formats are considered, we assume for simplicity that each polarization can be treated as an independent parallel channel. Under this assumption, and for the modulation formats studied in this paper (PM-16QAM, PM-64QAM, and PM-256QAM), the system under analysis can be reduced to a single-polarization (2D) one.      
At the transmitter, a CM encoder encodes a stream of $N_b$ information bits $\boldsymbol{B}^{N_b}=[B_1,B_2,\ldots,B_{N_b}]$ into a sequence of $N_s$ symbols $\boldsymbol{X}^{N_s}=[X_1,X_2,\ldots,X_{N_s}]$, each drawn from a set of $M$ complex values $\mathcal{S}=\{s_1,s_2,...,s_M\}$, where $M$ is a power of 2.\footnote{Throughout this paper, boldface uppercase variables (e.g., $\boldsymbol{X}^N$) denote random vectors where the superscript indicates the size of the vector. Calligraphic letters (e.g., $\mathcal{S}$) represent sets.} The rate at which this operation is performed (in bits per symbol) is therefore given by 
\begin{equation}
R=\frac{N_b}{N_s}.
\end{equation}   
In our analysis, we will only consider the case where the symbols $X_n$ forming a codeword $\boldsymbol{X}^{N_s}$ are independent, identically distributed (i.i.d.) random variables with equal probability $1/M$.\footnote{However, once a codebook is selected, symbols within codewords will appear as statistically dependent.}

Although all CM encoders are inherently nonbinary encoders, the encoding process described above can be implemented in two different ways, as shown in Fig.~\ref{fig:CM Encoder}. In the first implementation, shown in the top part of Fig.~\ref{fig:CM Encoder}, the sequence of information bits is encoded using a binary FEC code and subsequently a memoryless mapper $\Phi$ is used to convert blocks of $\log_2{M}$ bits into symbols of the constellation $\mathcal{S}$.\footnote{Throughout the paper, it is assumed that the mapping is done via the binary reflected Gray code\cite{Gray1953,Agrell2004}.} This implementation is naturally associated with CM decoders based on a demapper and a binary FEC decoder.  
The second implementation is shown in the bottom part of Fig.~\ref{fig:CM Encoder}, where bits are first mapped into a sequence of nonbinary information symbols, which are then mapped into sequences of nonbinary coded symbols by a nonbinary FEC encoder \cite{Schmalen2016}. In this case, the decoding can be performed by a nonbinary FEC decoder.

In this paper, we do not consider cases where symbols are not uniformly distributed, i.e., when a probabilistic shaping on $\mathcal{S}$ is performed \cite{Forney1984,Kschischang1993,Fischer2005,Yankov2014,Buchali2016,Fehenberger2016}. Moreover, throughout this paper, we focus our attention on high SEs ($>$2 bits/sym/polarization), and thus the constellation $\mathcal{S}$ is assumed to be a square $M$QAM constellation where $M\in\{16,64,256\}$.        

The symbols $X_n$ are mapped, one every $T_s$ seconds, onto a set of waveforms by a (real) pulse shaper $p(t)$, generating the complex signal 
\begin{equation}
s(t)=\sum_{n=1}^{N_\text{s}} X_{n}p(t-nT_s).
\label{eq:pulse_shaping}  
\end{equation}
The signal $s(t)$ propagates through $N_{\text{sp}}$ spans of optical fiber (see Fig.~\ref{fig:SystemSchematic}), optically amplified at the end of each span by an erbium-doped fiber amplifier (EDFA). At the end of the fiber link, the signal is detected by an optical receiver. As shown in Fig.~\ref{fig:SystemSchematic}, the first part of the receiver includes an equalizer and a matched filter (MF), which are assumed to be operating on the continuous-time received waveform $r(t)$.\footnote{The equalizer typically operates in the digital domain, but for a large enough sampling rate, the two representations are equivalent.} The equalizer performs a compensation of the most significant fiber channel impairments, either the linear ones only, as in the case of EDC, or both linear and nonlinear, as with DBP. 
The equalized (but noisy) waveform $y(t)$ represents the input of the detection stage and can be therefore effectively considered as the output of the so-called \textit{waveform channel}\cite[Sec.~2.4]{Agrell2016}. Such a channel is formed by the cascade of the physical channel and the equalization block at the receiver, as shown in Fig.~\ref{fig:SystemSchematic}. The physical channel (i.e., fiber spans and amplifiers), also referred to as nonlinear Schr\"{o}dinger channel in \cite{Kramer2015}, is described by the nonlinear Schr\"{o}dinger equation \cite[Sec.~2.3]{Agrawal2001}.

The receiver estimates the transmitted bits based on the set of observations $\boldsymbol{Y}^{N_s}$ that are extracted from the signal $y(t)$, using an MF matched to the transmitted pulse $p(t)$
\begin{equation}\label{eq:MF}
Y_n=\int_{-\infty}^{+\infty}y(\tau)p(\tau-nT_s)d\tau.
\end{equation} 
As shown in \cite{Liga2015,Irukulapati2014}, \eqref{eq:MF} does not necessarily represent the optimum way to reduce this particular waveform channel to a discrete-time one. However, the focus of this work is on the performance of CM encoder and decoder blocks, operating on the input and output of the \emph{discrete-time channel}, regardless of the suboptimality of the observations $\boldsymbol{Y}^{N_s}$.
  
In the following section, we will discuss AIRs of the four decoding strategies shown in Fig.~\ref{fig:Decoders}, representing different implementations of the CM decoder. The importance of these structures lies in the fact that they cover all main options employing a memoryless demapper. Each BW configuration (see Figs.~\ref{fig:Decoders}(b) and (d)) is characterized by a CM decoder formed by two blocks: a memoryless demapper and a binary FEC decoder. The SW strategies (see Figs.~\ref{fig:Decoders}(a) and (c)) are instead characterized by the adoption of a nonbinary decoder operating directly on symbol level metrics derived from the samples $Y_n$.
Each of the HD schemes (see Figs.~\ref{fig:Decoders}(c) and (d)) operates a symbol/bit level decision before the FEC decoder, which as a result operates on discrete quantities (\textit{hard information}). In the SD case (see Figs.~\ref{fig:Decoders}(a) and (b)), the decoder instead produces codeword estimates based on BW or SW log-likelihood (LL) values\footnote{For the binary case, LL \emph{ratios} are typically preferred for implementation reasons.}, which are distributed on a continuous range of values (\textit{soft information}). 

\section{AIRs for CM Systems}\label{sec3}
\subsection{Information-theoretic Preliminaries}\label{sec:inf_theory}
Consider an \textit{information stable}, discrete-time channel with memory \cite{Verdu1994}, characterized by the sequence of probability density functions (PDFs)\footnote{Throughout this paper, $p_{\boldsymbol{Y}|\boldsymbol{X}}(\boldsymbol{y}|\boldsymbol{x})$ denotes a joint conditional PDF for the random vectors $\boldsymbol{Y}$ and $\boldsymbol{X}$, whereas a marginal joint PDF is denoted by $p_{\boldsymbol{X}}(\boldsymbol{x})$.}
\begin{equation}
p_{\boldsymbol{Y}^N|\boldsymbol{X}^N}(\boldsymbol{y}^N|\boldsymbol{x}^N), \qquad N=1,2,\ldots  
\label{eq:ChLaw}
\end{equation}
The maximum rate at which reliable transmission over such a channel is possible is defined by the capacity \cite[eq.~(1.2)]{Verdu1994}:
\begin{equation}\label{eq:capacity}
C=\lim_{N\to \infty}\sup_{p_{\boldsymbol{X}^N}}\frac{1}{N}I(\boldsymbol{Y}^N;\boldsymbol{X}^N)
\end{equation}
where $p_{\boldsymbol{X}^N}$ is the joint PDF of the sequence $\boldsymbol{X}^N$ under a given power constraint.
When $p_{\boldsymbol{X}^N}$ is fixed, the quantity
\begin{equation}
 I(\boldsymbol{X}^N;\boldsymbol{Y}^N)=\mathbb{E}\left[\log_2{\frac{p_{\boldsymbol{Y}^N|\boldsymbol{X}^N}(\boldsymbol{Y}^N|\boldsymbol{X}^N)}{p_{\boldsymbol{Y}^N}(\boldsymbol{Y}^N)}}\right] 
\label{eq:MI2}
\end{equation}
in \eqref{eq:capacity} is the MI between the two sequences of symbols $\boldsymbol{X}^N$ and $\boldsymbol{Y}^N$, and
\begin{equation}
I_{\text{mem}}=\lim_{N\to \infty}\frac{1}{N} I(\boldsymbol{X}^N;\boldsymbol{Y}^N)
\label{eq:MI1}
\end{equation}
is the average \textit{per-symbol} MI rate \cite{Gallager1968,Secondini2013}, which has a meaning of channel MI. For a fixed $N$, \eqref{eq:MI1} represents the maximum AIR for the channel in \eqref{eq:ChLaw}, and can be achieved by a CM encoder generating codewords $\boldsymbol{X}^{N_{s}}$ according to $p_{\boldsymbol{X}^N}$, used along with an optimum decoder.\footnote{The channel can be seen as \emph{block-wise memoryless}, and thus, codewords should be constructed using blocks of $N$ symbols drawn using independently from $p_{\boldsymbol{X}^N}$.} Such a decoder uses the channel observations $\boldsymbol{y}^{N_s}$ to produce codeword estimates $\hat{\boldsymbol{X}}{}^{N_s}$ based on the rule
\begin{equation}\label{ML_Decoder}
\hat{\boldsymbol{X}}{}^{N_s}=\argmax_{\boldsymbol{x}^{N_s}\in \mathcal{S}^{N_s}}p_{\boldsymbol{Y}^{N_s}|\boldsymbol{X}^{N_s}}(\boldsymbol{y}^{N_s}|\boldsymbol{x}^{N_s})
\end{equation} 
where the codeword likelihood $p_{\boldsymbol{Y}^{N_s}|\boldsymbol{X}^{N_s}}$ is calculated based on the knowledge of the channel law \eqref{eq:ChLaw}. 

The expression of the channel law \eqref{eq:ChLaw}, for $N$ large enough to account for the channel memory, remains so far unknown for the optical fiber channel despite previous attempts to derive approximated \cite{Dar2014,Marsella2014} or heuristic \cite{Agrell2014} analytical expressions. On the other hand, brute-force numerical approaches appear prohibitive.        
An immediate consequence is that the exact channel MI for a given modulation format cannot be calculated. The second consequence is that the optimum receiver potentially achieving a rate $R=I_{\text{mem}}$ cannot be designed. However, using the mismatched decoder approach, it is still possible to calculate nontrivial AIRs for the optical fiber channel in Fig.~\ref{fig:SystemSchematic}, when suboptimal but practically realizable CM encoders and decoders are used, such as the ones described in Section~\ref{sec2} (see Fig.~\ref{fig:Decoders}).

The method of the mismatched decoder to calculate AIRs for specific decoder structures originates from the works in \cite{Merhav1994}, later extended to channels with memory in \cite{Arnold2006} and recently applied to optical fiber systems in, e.g., \cite{Colavolpe2011, Secondini2013,Fehenberger2015}. This approach consists of replacing, in the calculation of the channel MI, the unknown channel law with an auxiliary one, obtaining a lower bound. Moreover, such a bound represents an AIR for a system using the optimum decoder for the auxiliary channel. The tightness of such a lower bound depends on how similar the auxiliary channel is to the actual one. On the other hand, no converse coding theorem is available for the bound obtained using a given auxiliary channel. In other words, even when a mismatched decoder is used, the estimated rate is not necessarily the maximum achievable rate. Counterexamples have been shown, e.g., in \cite{Martinez2015}.

Nevertheless the AIRs calculated via the mismatched decoder approach still represent an upper bound on the rates of most, if not all, coding schemes used in practice. Furthermore they are a strong predictor of the post-FEC BER of such schemes, as shown in \cite{Leven2011,Schmalen2012,Alvarado2015,Schmalen2016}.
\subsection{AIRs for SD CM Decoders}\label{sec:SD_CM_Dec}
Since each of the CM decoders presented in Section~\ref{sec2} neglects the memory of the channel in \eqref{eq:ChLaw}, a first decoding \textit{mismatch} is introduced. In what follows, we will discuss this mismatch using the SD-SW case (see Fig.~\ref{fig:Decoders}(a)) as a representative example of all other CM decoders.

For the SD-SW, the nonbinary decoder requires SW likelihoods $p_{Y_n|X_n}$, with $n=1,2,\ldots,N$.  These $N$ PDFs can be derived for each $n$ by marginalizing the joint PDF in \eqref{eq:ChLaw}. For simplicity, however, practical implementations use a single PDF across the block of $N$ symbols. We choose the PDF in the middle of the observation block, i.e., at time instant $n = \hat{n} = \lceil N/2 \rceil$. The marginalization of \eqref{eq:ChLaw} in this case gives
\begin{equation}
p_{Y_{\hat{n}}|X_{\hat{n}}}(y_{\hat{n}}|x_{\hat{n}})=\int_{\mathbb{C}^{N-1}}p_{\boldsymbol{Y}^N|X_{\hat{n}}}(\boldsymbol{y}^N|x_{\hat{n}}) d\tilde{\boldsymbol{y}}^{N-1}
\label{eq:MemLessChLaw1}
\end{equation} 
where $\mathbb{C}$ denotes the complex field,
$\tilde{\boldsymbol{y}}^{N-1}\triangleq[y_{1},\ldots,y_{\hat{n}-1},y_{\hat{n}+1},\ldots,y_{N}]$, and the conditional PDF $p_{\boldsymbol{Y}^N|X_{\hat{n}}}$ in \eqref{eq:MemLessChLaw1} can be expressed as
\begin{equation}
p_{\boldsymbol{Y}^N|X_{\hat{n}}}(\boldsymbol{y}^N|x_{\hat{n}})=\frac{1}{M^{N-1}}\sum_{\tilde{\boldsymbol{x}}^{N-1}\in\mathcal{S}^{N-1}}p_{\boldsymbol{Y}^N|\boldsymbol{X}^N}(\boldsymbol{y}^N|\boldsymbol{x}^{N})
\label{eq:MemLessChLaw2}
\end{equation}
where $\tilde{\boldsymbol{x}}^{N-1}\triangleq[x_{1},\ldots,x_{\hat{n}-1},x_{\hat{n}+1},\ldots,x_{N}]$.

The choice for the single PDF to be the one in the middle of the observation block is arbitrary. However, this choice is justified by the fact that $p_{Y_{\hat{n}}|X_{\hat{n}}}(y_{\hat{n}}|x_{\hat{n}})$ will be a good approximation of all other PDFs $p_{Y_n|X_n}(y_n|x_n)$ with $n=1,2,\ldots, N$ when $N$ is large.

The demapper is then assuming a channel that is stationary across the block of $N$ symbols.\footnote{Here we refer to \emph{wide-sense} stationarity \cite[Sec.~3.6.1]{Gallager2014}.} This channel is fully determined by a PDF $p_{Y|X}(y|x)$ defined as
\begin{align}
\label{eq:MemLessChLawStat}
p_{Y|X}(y|x) &\triangleq p_{Y_{\hat{n}}|X_{\hat{n}}}(y|x)=p_{Y_n|X_n}(y|x)
\end{align}
with $n=1,2,\ldots, N$. When i.i.d. symbols are transmitted, the MI for this auxiliary memoryless channel is given by
\begin{equation}
I_{\text{\tiny{SD-SW}}}=\frac{1}{M}\sum_{i=1}^{M}\int_{\mathbb{C}}p_{Y|X}(y|s_i)\log_2{\frac{p_{Y|X}(y|s_i)}{p_{Y}(y)}}dy.
\label{eq:SD-SW_MI}
\end{equation}

The SD-SW MI in \eqref{eq:SD-SW_MI} is an AIR for the SD-SW decoder structure in Fig.~\ref{fig:Decoders}(a), where the demapper computes LLs $\log p_{Y|X}(y|x)$, and the FEC decoder estimates each transmitted codeword using \eqref{ML_Decoder} with a codeword likelihood given by
\begin{equation}
p_{\boldsymbol{Y}^{N_s}|\boldsymbol{X}^{N_s}}(\boldsymbol{y}^{N_s}|\boldsymbol{x}^{N_s}) =\prod_{n=1}^{N_s}p_{Y|X}(y_n|x_n). 
\end{equation}

In most cases, the channel law $p_{\boldsymbol{Y}^N|\boldsymbol{X}^N}$ is unknown and therefore $p_{Y|X}(y|x)$ is not available in closed form to the receiver. Also, numerical estimations of $p_{Y|X}(y|x)$ are often prohibitive. As a result, practical implementations not only ignore the memory of the channel (first mismatch), but also make an a priori assumption on the PDF $p_{Y|X}(y|x)$. This assumption introduces a \emph{second} mismatch, which we discuss in what follows.

Most receivers assume a circularly symmetric Gaussian distribution for \eqref{eq:MemLessChLawStat}. In this case, an AIR is given by \cite[eq.~(2)]{Fehenberger2015} 
\begin{equation}
\tilde{I}_{\text{\tiny{SD-SW}}}=\frac{1}{M}\sum_{i=1}^{M}\int_{\mathbb{C}}p_{Y|X}(y|s_i)\log_2{\frac{q_{Y|X}(y|s_i)}{q_{Y}(y)}}dy
\label{eq:Mismatched_SD-SW_MI}
\end{equation}
where 
\begin{equation}
q_{Y|X}(y|x)=\frac{1}{\pi\sigma^2}\exp\left(-\frac{|y-x|^2}{\sigma^2}\right)
\label{eq:aux}
\end{equation}
represents the auxiliary Gaussian channel with complex noise variance $\sigma^2$, which accounts for the contributions of both ASE and nonlinear distortions.

As shown in \cite{Poggiolini2014,Carena2012}, the marginal PDF for the optical fiber channel is in most practical cases well approximated by a circularly symmetric Gaussian distribution.\footnote{A deviation from a circularly symmetric Gaussian PDF can be observed, e.g., in the following cases: amplification schemes different from EDFA (such as Raman amplifiers)\cite{Dar2014}, dispersion-managed links (see for instance \cite{Secondini2013}), and for very high transmitted powers.} Therefore, as pointed out in \cite{Fehenberger2015}, we generally have
\begin{equation}
\tilde{I}_{\text{\tiny{SD-SW}}}\approx I_{\text{\tiny{SD-SW}}}.
\end{equation} 
In this case, as we will discuss in Sec.~\ref{sec4}, the AIRs of SD-SW decoders can be quite accurately estimated using the MI expression for the AWGN channel and the effective signal-to-noise ratio (SNR) at the MF output 
\begin{equation}
\text{SNR}=\frac{\mathbb{E}\left[|X|^2\right]}{\sigma^2}.
\label{eq:SNR}
\end{equation}

In the SD-BW implementation (see Fig.~\ref{fig:Decoders}(b)), for each received symbol $Y$ the demapper generates $\log_2{M}$ BW LLs \cite{Alvarado2015}, \cite[Ch.~3]{Szczecinski2015bit}. These LLs are usually obtained assuming no statistical dependence between bits belonging to the same transmitted symbol. When such LLs are calculated based on a memoryless channel law $p_{Y|X}(y|x)$, the relevant quantity for the coded performance is the GMI \cite[eq.~(4.54)]{Szczecinski2015bit}, \cite[eq.~(24)]{Alvarado2015}
\begin{equation}
I_{\text{\tiny{SD-BW}}}=\sum_{k=1}^{\log_2{M}}I(B_k;Y)
\label{eq:GMI}
\end{equation}
where $B_{k}$ denotes the $k$-th bit of $X$ and $I(B_{k};Y)$ denotes the MI between transmitted bits and received symbols.

When the LLs are calculated using the auxiliary channel in \eqref{eq:aux} instead of the true channel, the GMI is lower-bounded by
\begin{align}
\tilde{I}_{\text{\tiny{SD-BW}}}&=\frac{1}{M}\sum_{k=1}^{\log_2{M}}\sum_{b\in\{0,1\}}\sum_{i\in\mathcal{I}_k^b}\int_{\mathbb{C}}p_{Y|X}(y|s_i)g_{k,b}(y)dy
\label{eq:MisGMI}
\end{align} 
where $\mathcal{I}_k^b$ is the subset of indices of the constellation $\mathcal{S}$ having the $k$-th bit equal to $b\in\{0,1\}$ and 
\begin{equation}
g_{k,b}(y)\triangleq\log_2{\frac{\sum_{j\in\mathcal{I}_k^b} q_{Y|X}(y|s_j)}{\frac{1}{2}\sum_{j=1}^{M}q_{Y|X}(y|s_j)}}.
\label{eq:intra_log}
\end{equation}
Similarly to the SD-SW case, for the optical fiber channel in Fig.~\ref{fig:SystemSchematic} we have $\tilde{I}_{\text{\tiny{SD-BW}}}\approx I_{\text{\tiny{SD-BW}}}$.

\subsection{AIRs for HD CM Decoders}\label{sec:HD_CM_Dec}
As illustrated in Figs.~\ref{fig:Decoders}(c) and (d), the HD decoders are preceded by a threshold device casting the channel samples $\boldsymbol{Y}^{N_s}$ into a discrete set of values. In the SW case (Fig.~\ref{fig:Decoders}(c)), such a device provides a sequence of hard SW estimates $\hat{\boldsymbol{X}}{}^{N_s}$ that are passed to a nonbinary decoder. 
The channel 
\begin{equation}
P_{\hat{\boldsymbol{X}}{}^N|\boldsymbol{X}^N}(\hat{\boldsymbol{x}}{}^N|\boldsymbol{x}^N)
\label{eq:HD-SWChLaw}
\end{equation}
will in general show memory across multiple symbols $\hat{X}_n$. However, in analogy with \eqref{eq:MemLessChLaw1}, we can replace \eqref{eq:HD-SWChLaw} with an equivalent memoryless channel defined by    
\begin{equation}
P_{\hat{X}|X}(x_j|x_i)\triangleq p_{ij} \;\;\; \text{for} \ \ i,j=1,2,...,M
\label{eq:HD-SW_Ch}
\end{equation}
where the $p_{ij}$ are the SW crossover probabilities.
Using the same argument on the channel memory used for the SD-SW case, the quantity
\begin{equation}
I_{\text{\tiny{HD-SW}}}=\frac{1}{M}\sum_{i=1}^{M}\sum_{j=1}^{M}p_{ij}\log_2{\frac{p_{ij}}{\frac{1}{M}\sum_{p=1}^{M}p_{pj}}}
\label{eq:HD-SW_MI}
\end{equation}
represents an AIR for the HD-SW CM decoder in Fig.~\ref{fig:Decoders}(c).\footnote{The rate $I_{\text{\tiny{HD-SW}}}$ in \eqref{eq:HD-SW_MI} is achievable with a nonbinary FEC decoder that is matched to the channel transition probabilities $p_{i,j}$, but not necessarily with a standard nonbinary FEC decoder based on minimizing the Hamming distance.}
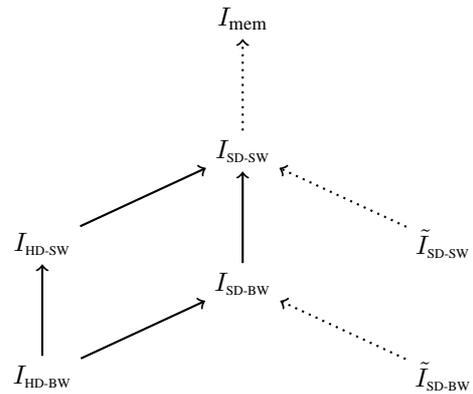
\begin{figure}[!t]
\begin{center}
\begin{tikzpicture}
\node[] at (0pt,0pt){};
\node[] at (0pt,135pt)(Imem){$I_{\text{mem}}$};
\node[] at (0pt,85pt) (SD-SW){$I_{\text{\tiny{SD-SW}}}$};
\node[] at (75pt,50pt) (Mis_SD-SW){$\tilde{I}_{\text{\tiny{SD-SW}}}$};
\node[] at (0pt,35pt) (SD-BW){$I_{\text{\tiny{SD-BW}}}$};
\node[] at (75pt,0pt) (Mis_SD-BW){$\tilde{I}_{\text{\tiny{SD-BW}}}$};
\node[] at (-75pt,50pt) (HD-SW){$I_{\text{\tiny{HD-SW}}}$};
\node[] at (-75pt,0pt) (HD-BW){$I_{\text{\tiny{HD-BW}}}$};
\draw[->,thick,dotted] (SD-SW)--(Imem);
\draw[->,thick,dotted] (Mis_SD-SW)--(SD-SW);
\draw[->,thick,dotted] (Mis_SD-BW)--(SD-BW);
\draw[->,thick] (SD-BW)--(SD-SW);
\draw[->,thick] (HD-BW)--(HD-SW);
\draw[->,thick] (HD-SW)--(SD-SW);
\draw[->,thick] (HD-BW)--(SD-BW);
\end{tikzpicture}
%
\caption{Graph showing relationships between the information-theoretic quantities presented in this paper. Lines between nodes indicate an inequality, where the arrows point towards the upper bound. Dotted arrows indicate inequalities which become equalities for the AWGN channel.}
\label{fig:AIR_rel}
\end{center}
\end{figure}
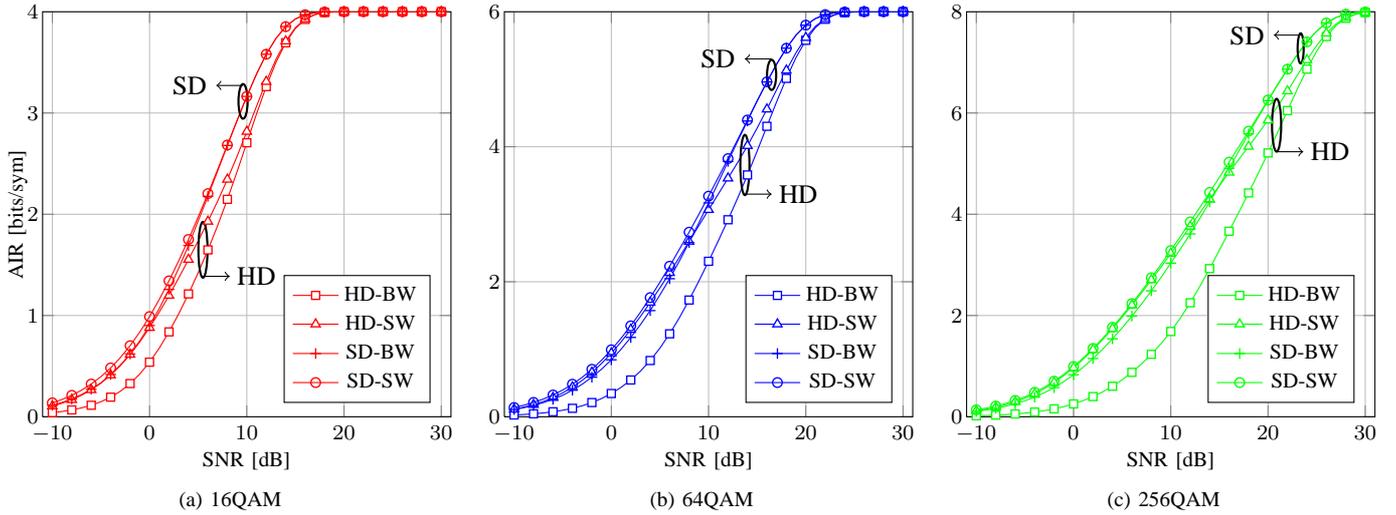
\begin{figure*}[!t]
\hspace{-.45cm}
\setlength{\tabcolsep}{0.0001cm}
\begin{tabular}{ccc}
\subfloat[16QAM]{
\begin{tikzpicture}
\begin{axis}[
	legend style={at={(0.95,0.35)},font=\footnotesize,
	legend style={row sep=0.05pt}},
	ticklabel style = {font=\footnotesize},
	width=0.785\columnwidth,
	height=0.785\columnwidth,
	grid=both,
	xmin=-11,xmax=31,
	ymin=0,ymax=4,  
	xlabel={\footnotesize SNR [dB]},
	xlabel shift= -3 pt,
	ylabel={\footnotesize AIR [bits/sym]},
	ylabel shift = -5 pt,
	yticklabel shift = -2 pt]
\addlegendentry{HD-BW};
\addplot[color=red,thin, mark=square*,mark size=1.5,mark options={fill=white,line width=0.5},mark repeat={2}] file {Figures_data/AWGN_16QAM_HD-BW.txt};
\addlegendentry{HD-SW};
\addplot[color=red,thin, mark=triangle*,mark size=2,mark options={fill=white,line width=0.5},mark repeat={2}] file {Figures_data/AWGN_16QAM_HD-SW.txt};
\addlegendentry{SD-BW};
\addplot[color=red,thin, mark=+,mark size=2,mark options={fill=white,line width=0.5},mark repeat={2}] file {Figures_data/AWGN_16QAM_SD-BW.txt};
\addlegendentry{SD-SW};
\addplot[color=red,thin, mark=o, mark size=1.8,mark repeat={2},mark options={fill=white,line width=0.5}] file {Figures_data/AWGN_16QAM_SD-SW.txt};
\draw[thick] (60pt,63pt) ellipse (0.06cm and 0.37cm);
\draw[->] (60pt,53pt) -- (70pt,53pt);
\node at (80pt,53pt) {HD};
\draw[thick] (75pt,119pt) ellipse (0.06cm and 0.23cm);
\draw[->] (75pt,125pt) -- (65pt,125pt);
\node at (55pt,125pt) {SD};
\end{axis}
\end{tikzpicture}\label{subfig:16QAM}
} &
\subfloat[64QAM]{
\begin{tikzpicture}
\begin{axis}[
	legend style={at={(0.95,0.35)},font=\footnotesize,
	legend style={row sep=0.05pt}},
	ticklabel style = {font=\footnotesize},
	width=0.785\columnwidth,
	height=0.785\columnwidth,
	grid=both,
	xmin=-11,xmax=31,
	ymin=0,ymax=6,  
	xlabel={\footnotesize SNR [dB]},
	xlabel shift= -3 pt,
	ylabel shift = -5 pt,
	yticklabel shift = -2 pt]
\addlegendentry{HD-BW};
\addplot[color=blue,thin, mark=square*,mark size=1.5,mark options={fill=white,line width=0.5},mark repeat={2}] file {Figures_data/AWGN_64QAM_HD-BW.txt};
\addlegendentry{HD-SW};
\addplot[color=blue,thin, mark=triangle*,mark size=2,mark options={fill=white,line width=0.5},mark repeat={2}] file {Figures_data/AWGN_64QAM_HD-SW.txt};
\addlegendentry{SD-BW};
\addplot[color=blue,thin, mark=+,mark size=2,mark options={fill=white,line width=0.5},mark repeat={2}] file {Figures_data/AWGN_64QAM_SD-BW.txt};
\addlegendentry{SD-SW};
\addplot[color=blue,thin, mark=o, mark size=1.8,mark options={fill=white,line width=0.5},mark repeat={2}] file {Figures_data/AWGN_64QAM_SD-SW.txt};
\draw[thick] (90pt,95pt) ellipse (0.06cm and 0.4cm);
\draw[->] (90pt,84pt) -- (100pt,84pt);
\node at (110pt,84pt) {HD};
\draw[thick] (100pt,129pt) ellipse (0.06cm and 0.2cm);
\draw[->] (100pt,135pt) -- (90pt,135pt);
\node at (80pt,135pt) {SD};
\end{axis}
\end{tikzpicture}\label{subfig:64QAM}
} &
\subfloat[256QAM]{
\begin{tikzpicture}
\begin{axis}[
	legend style={at={(0.95,0.35)},font=\footnotesize,
	legend style={row sep=0.05pt}},
	ticklabel style = {font=\footnotesize},
	width=0.785\columnwidth,
	height=0.785\columnwidth,
	grid=both,
	xmin=-11,xmax=31,
	ymin=0,ymax=8,  
	xlabel={\footnotesize SNR [dB]},
	xlabel shift= -3 pt,
	ylabel shift = -5 pt,
	yticklabel shift = -2 pt]
\addlegendentry{HD-BW};
\addplot[color=green,thin, mark=square*,mark size=1.5,mark options={fill=white,line width=0.5},mark repeat={2}] file {Figures_data/AWGN_256QAM_HD-BW.txt};
\addlegendentry{HD-SW};
\addplot[color=green,thin, mark=triangle*,mark size=2,mark options={fill=white,line width=0.5},mark repeat={2}] file {Figures_data/AWGN_256QAM_HD-SW.txt};
\addlegendentry{SD-BW};
\addplot[color=green,thin, mark=+,mark size=2,mark options={fill=white,line width=0.5},mark repeat={2}] file {Figures_data/AWGN_256QAM_SD-BW.txt};
\addlegendentry{SD-SW};
\addplot[color=green,thin, mark=o, mark size=1.8,mark options={fill=white,line width=0.5},mark repeat={2}] file {Figures_data/AWGN_256QAM_SD-SW.txt};
\draw[thick] (116pt,110pt) ellipse (0.06cm and 0.35cm);
\draw[->] (116pt,100pt) -- (126pt,100pt);
\node at (136pt,100pt) {HD};
\draw[thick] (125pt,139pt) ellipse (0.04cm and 0.2cm);
\draw[->] (125pt,144pt) -- (115pt,144pt);
\node at (105pt,144pt) {SD};
\end{axis} 
\end{tikzpicture}\label{subfig:256QAM}
} 
\end{tabular}
\caption{AIRs vs. SNR for different modulation formats for the AWGN channel.} 
\label{fig:MI_AWGN}
\end{figure*} 
When the HD decoder structure is preserved but a \emph{binary} decoder is instead used (Fig.~\ref{fig:Decoders}(d)), the threshold device needs to be followed by a symbol-to-bit demapper producing a sequence of pre-FEC bits estimates $\hat{\boldsymbol{B}}{}^{N_b}$. Again, although the resulting binary channel might show memory, the HD FEC decoder typically neglects it and the most likely codeword is calculated based on each single detected bits. The marginal channel law $P_{\hat{B}|B}(\hat{b}|b)$ is in this case represented by a binary symmetric auxiliary channel\footnote{The bit position within each symbol is here disregarded and an average conditional PDF is considered.}
\begin{equation}
P_{\hat{B}|B}(\hat{b}|b)=\left\{
\begin{array}{ll}
      p & \text{for} \ \ \hat{b}\neq b \\
      1-p & \text{for} \ \  \hat{b}=b   \\    
\end{array}
\right.
\label{eq:HDBinChLaw}
\end{equation}  
where $p$ corresponds to the average pre-FEC BER (BW crossover probability). 
The quantity
\begin{equation}
I_{\text{\tiny{HD-BW}}}=m[1+p\log_2p+(1-p)\log_2(1-p)]
\label{eq:HD-BW_MI}
\end{equation}
then represents an AIR for an HD-BW CM decoder in Fig.~\ref{fig:Decoders}(d). 
 
\subsection{Relationships Between AIRs}
The relationships between the above discussed 
AIRs are summarized by means of the graph in  Fig.~\ref{fig:AIR_rel}. Nodes that are connected in the graph indicate the existence of an inequality between the quantities in each of the nodes. The direction of the arrows show which quantity is upper-bounding the other.

For any given input distribution, the rate $I_{\text{mem}}$ upper-bounds all other quantities. In particular we have
\begin{equation}
I_{\text{mem}}\geq I_{\text{\tiny{SD-SW}}}\geq \tilde{I}_{\text{\tiny{SD-SW}}}.
\end{equation}    
where the first inequality can be proven using the chain rule of the MI (see \cite{Wolfowitz1967},\cite[Sec.~IV]{Essiambre2010},\cite[Sec.~2.5.2]{Cover2006}). 
The second inequality instead reflects the additional mismatch caused by a memoryless demapper based on \eqref{eq:aux} and not on \eqref{eq:MemLessChLaw1}. The proof of this inequality follows from the definitions \eqref{eq:SD-SW_MI} and \eqref{eq:Mismatched_SD-SW_MI} and is given in \cite[Sec.~VI]{Arnold2006}. 
Due to the assumption of independent bits within each transmitted symbol in the calculation of \eqref{eq:GMI}, it can also be shown that \cite[Sec.~4.4]{Szczecinski2015bit} 
\begin{equation}
I_{\text{\tiny{SD-SW}}}\geq I_{\text{\tiny{SD-BW}}}\geq \tilde{I}_{\text{\tiny{SD-BW}}}.
\label{eq:ineq1}
\end{equation} 
Again, the second inequality reflects the loss of information of a mismatched demapper calculating BW LLs based on \eqref{eq:aux} rather than on \eqref{eq:MemLessChLaw1}.  

Due to the data-processing inequality \cite[Sec.~2.4]{Cover2006} and the mismatch of the illustrated HD decoders to the potential channel memory, we have
\begin{align}
I_{\text{\tiny{SD-SW}}}\geq I_{\text{\tiny{HD-SW}}},\\
I_{\text{\tiny{SD-BW}}}\geq I_{\text{\tiny{HD-BW}}}.
\label{eq:ineq2}
\end{align}
Finally, similarly to the SD case, we have
\begin{equation}
I_{\text{\tiny{HD-SW}}}\geq I_{\text{\tiny{HD-BW}}}.
\label{eq:ineq3}
\end{equation} 

In general, nothing can be said on the relationship between $I_{\text{\tiny{SD-BW}}}$ and $I_{\text{\tiny{HD-SW}}}$.
Also, no systematic inequality holds between the \textit{mismatched} versions of the SD AIRs ($\tilde{I}_{\text{\tiny{SD-SW}}}$, $\tilde{I}_{\text{\tiny{SD-BW}}}$) and the HD AIRs ($I_{\text{\tiny{HD-SW}}}$, $I_{\text{\tiny{HD-BW}}}$). However, as already discussed in Section~\ref{sec:SD_CM_Dec}, for the optical fiber channel the mismatched AIRs are expected to be very close to the AIRs obtained with perfect knowledge of the channel marginal PDF in \eqref{eq:MemLessChLaw1}. 
 
When the channel is indeed AWGN, clearly
\begin{align}
I_{\text{mem}}&=I_{\text{\tiny{SD-SW}}}=\tilde{I}_{\text{\tiny{SD-SW}}},\\
I_{\text{\tiny{SD-BW}}}&=\tilde{I}_{\text{\tiny{SD-BW}}}
\end{align}
In this case, as illustrated in \ref{fig:AIR_rel}, $I_{\text{\tiny{SD-SW}}}$ and $I_{\text{\tiny{HD-SW}}}$ are the maximum AIR for SD-SW and HD-SW decoders, respectively \cite{Shannon1948}, since each demapper is matched to the channel.\footnote{In the HD-SW case, the channel seen by the nonbinary FEC decoder is the one in \eqref{eq:HD-SW_Ch}.} Conversely, for BW decoders, rates higher than $I_{\text{\tiny{SD-BW}}}$ and $I_{\text{\tiny{HD-BW}}}$ are still  possible (see, e.g., \cite{Martinez2015}).

In order to better illustrate the relationships discussed above, the four AIRs in \eqref{eq:SD-SW_MI}, \eqref{eq:GMI}, \eqref{eq:HD-SW_MI}, and \eqref{eq:HD-BW_MI} were calculated for the AWGN channel. In Fig.~\ref{fig:MI_AWGN}, $I_{\text{\tiny{SD-SW}}}$, $I_{\text{\tiny{SD-BW}}}$, $I_{\text{\tiny{HD-SW}}}$, and $I_{\text{\tiny{HD-BW}}}$ are shown vs. the SNR in \eqref{eq:SNR} for the three $M$QAM formats analyzed in this paper: 16QAM, 64QAM, and 256QAM. For 16QAM, the HD AIRs are below both of the SD AIRs. It  should be noted that for SD decoders, a negligible penalty is incurred by using a BW structure. As the modulation order is increased, and for low enough SNR values, it can be observed that the HD-SW AIRs match or exceed the SD-BW AIRs. Also, in this regime, the performance of these two decoders are comparable to the SD-SW one. This behaviour is clearer for a 256QAM modulation format, where a more significant penalty is incurred by using BW demapping in an SD CM decoder, whereas the HD-SW structure performs as well as the SD counterpart. When the modulation format cardinality increases, an HD-BW decoder incurs, in general, significant penalties in AIR.  Finally, the inequalities in \eqref{eq:ineq1}--\eqref{eq:ineq3} can be seen to hold for all modulation formats shown, as expected.  
%
%
%

\section{Numerical Results}\label{sec4}
\subsection{Numerical Setup}
In this section, numerical results based on split-step Fourier (SSF) simulations of optical fiber transmission are presented. As shown in Fig.~\ref{fig:SystemSchematic}, the simulated system consists of an optical fiber link comprising multiple standard single mode (SMF) fiber spans (parameters shown in Table~\ref{tab:syspar}), amplified, at the end of each span, by an EDFA which compensates for the span loss. At the transmitter, after the CM encoder, PM square $M$QAM formats (PM-16QAM, PM-64QAM, PM-256QAM) were modulated using a root raised cosine (RRC) filter $p(t)$. For each polarization of each WDM channel, independent sequences of $2^{18}$ symbols were transmitted.
The fiber propagation was simulated by numerically solving the Manakov equation through the SSF method. In order to obtain ideal equalization performance, the sampling rate at which the equalizer was operated was the same as the fiber propagation simulation (512 GSa/s). 

After the MF (see Fig.~\ref{fig:SystemSchematic}) and sampling at 1 Sa/sym, AIRs calculations were performed based on the schemes shown in Fig.~\ref{fig:Decoders}. In particular, we used \eqref{eq:Mismatched_SD-SW_MI}--\eqref{eq:aux}, \eqref{eq:MisGMI}--\eqref{eq:intra_log}, \eqref{eq:HD-SW_MI}, and \eqref{eq:HD-BW_MI} to evaluate $\tilde{I}_{\text{\tiny{SD-SW}}}$, $\tilde{I}_{\text{\tiny{SD-BW}}}$, $I_{\text{\tiny{HD-SW}}}$, and $I_{\text{\tiny{HD-BW}}}$, respectively.
For the calculation of $\tilde{I}_{\text{\tiny{SD-SW}}}$ and $\tilde{I}_{\text{\tiny{SD-BW}}}$ in \eqref{eq:Mismatched_SD-SW_MI} and \eqref{eq:MisGMI}, Monte-Carlo integration was performed, using the $2^{18}$ channel samples (transmitted symbols) to estimate the variance $\sigma^2$ of $q_{Y|X}(y|x)$. However, we found that $\tilde{I}_{\text{\tiny{SD-SW}}}\approx I_{\text{\tiny{SD-SW}}}$, when in $I_{\text{\tiny{SD-SW}}}$ $p_{Y|X}(y|x)$ is replaced by $q_{Y|X}(y|x)$, further confirming the Gaussianity of $p_{Y|X}(y|x)$. In order to calculate $I_{\text{\tiny{HD-SW}}}$ and $I_{\text{\tiny{HD-BW}}}$, a Monte-Carlo estimation of the probabilities $p_{ij}$ and $p$ was performed using the pairs of sequences $(\boldsymbol{X}^{N_s}$, $\hat{\boldsymbol{X}}{}^{N_s})$ and $(\boldsymbol{B}^{N_b}$, $\hat{\boldsymbol{B}}{}^{N_b})$, respectively.
 
\subsection{Optical Fiber AIRs}
\begin{figure*}[!htbp]
\hspace{-.45cm}
\setlength{\tabcolsep}{0.001cm}
\begin{tabular}{ccc}
\subfloat[PM-16QAM]{
\label{subfig:EDC_16QAM}
\begin{tikzpicture}
\begin{axis}[
	legend style={at={(0.98,0.98)},font=\footnotesize,
	legend style={row sep=0.05pt}},
	ticklabel style = {font=\footnotesize},
	xtick={1000,2000,4000,6000, 8000, 10000, 12000},
xticklabels={1000,,4000, 6000, 8000, 10000,12000},
	width=0.785\columnwidth,
	height=0.76\columnwidth,
	grid=both,
	xmin=500,xmax=12700,
	ymax=8,  
	xlabel={\footnotesize Distance [km]},
	xlabel shift= -3 pt,
	ylabel={\footnotesize AIR [bits/sym]},
	ylabel shift = -5 pt,
	yticklabel shift = -2 pt]
\addlegendentry{HD-BW};
\addplot[color=red,thin, mark=square*,mark size=1.5,mark options={fill=white,line width=0.5}] file {Figures_data/EDC_only_16QAM_HD_BW_MI.txt};
\addlegendentry{HD-SW};
\addplot[color=red,thin, mark=triangle*,mark size=2,mark options={fill=white,line width=0.5}] file {Figures_data/EDC_only_16QAM_HD_SW_MI.txt};
\addlegendentry{SD-BW};
\addplot[color=red,thin, mark=+,mark size=2,mark options={fill=white,line width=0.5}] file {Figures_data/EDC_only_16QAM_SD_BW_MI.txt};
\addlegendentry{SD-SW};
\addplot[color=red,thin, mark=o, mark size=1.8,mark options={fill=white,line width=0.5}] file {Figures_data/EDC_only_16QAM_SD_SW_MI.txt};

\draw[thick] (90pt,47pt) ellipse (0.06cm and 0.3cm);
\draw[->] (90pt,39pt) -- (80pt,39pt);
\node at (70pt,39pt) {HD};
\draw[thick] (112pt,59pt) ellipse (0.06cm and 0.2cm);
\draw[->] (112pt,65pt) -- (122pt,65pt);
\node at (132pt,65pt) {SD};
\end{axis}
\end{tikzpicture}} &

\subfloat[PM-64QAM]{
\label{subfig:EDC_64QAM}
\begin{tikzpicture}
\begin{axis}[
	legend style={at={(0.98,0.98)},font=\footnotesize,
	legend style={row sep=0.05pt}},
	ticklabel style = {font=\footnotesize},
	xtick={1000, 2000, 4000, 6000, 8000, 10000, 12000},
xticklabels={1000,, 4000, 6000, 8000, 10000,12000},
	width=0.785\columnwidth,
	height=0.76\columnwidth,
	grid=both,
	xmin=500,xmax=12700,
	ymax=12,  
	xlabel={\footnotesize Distance [km]},
	xlabel shift= -3 pt,
	ylabel shift = -5 pt,
	yticklabel shift = -2 pt]
\addlegendentry{HD-BW};
\addplot[color=blue,thin, mark=square*,mark size=1.5,mark options={fill=white,line width=0.5}] file {Figures_data/EDC_only_64QAM_HD_BW_MI.txt};
\addlegendentry{HD-SW};
\addplot[color=blue,thin, mark=triangle*,mark size=2,mark options={fill=white,line width=0.5}] file {Figures_data/EDC_only_64QAM_HD_SW_MI.txt};
\addlegendentry{SD-BW};
\addplot[color=blue,thin, mark=+,mark size=2,mark options={fill=white,line width=0.5}] file {Figures_data/EDC_only_64QAM_SD_BW_MI.txt};
\addlegendentry{SD-SW};
\addplot[color=blue,thin, mark=o, mark size=1.8,mark options={fill=white,line width=0.5}] file {Figures_data/EDC_only_64QAM_SD_SW_MI.txt};
\addplot[color=red, mark=*,mark options={fill=red,line width=0.5}, mark size=2.2] coordinates {(10400,5.6)};

\draw[thick] (55pt,48pt) ellipse (0.06cm and 0.5cm);
\draw[->] (55pt,34pt) -- (45pt,34pt);
\node at (35pt,34pt) {HD};
\draw[thick] (67pt,60pt) ellipse (0.06cm and 0.2cm);
\draw[->] (67pt,65.5pt) -- (77pt,65.5pt);
\node at (87pt,65.5pt) {SD};
\end{axis}
\end{tikzpicture}}

\subfloat[PM-256QAM]{
\label{subfig:EDC_256QAM}
\begin{tikzpicture}
\begin{axis}[
	legend style={at={(0.98,0.98)},font=\footnotesize,
	legend style={row sep=0.05pt}},
	ticklabel style = {font=\footnotesize},
	xtick={1000, 2000, 4000, 6000, 8000, 10000, 12000},
xticklabels={1000, ,4000, 6000, 8000, 10000,12000},
	width=0.785\columnwidth,
	height=0.76\columnwidth,
	grid=both,
	xmin=500,xmax=12700,
	ymax=14,  
	xlabel={\footnotesize Distance [km]},
	xlabel shift= -3 pt,
	ylabel shift = -5 pt,
	yticklabel shift = -2 pt]
\addlegendentry{HD-BW};
\addplot[color=green,thin, mark=square*,mark size=1.5,mark options={fill=white,line width=0.5}] file {Figures_data/EDC_only_256QAM_HD_BW_MI.txt};
\addlegendentry{HD-SW};
\addplot[color=green,thin, mark=triangle*,mark size=2,mark options={fill=white,line width=0.5}] file {Figures_data/EDC_only_256QAM_HD_SW_MI.txt};
\addlegendentry{SD-BW};
\addplot[color=green,thin, mark=+,mark size=2,mark options={fill=white,line width=0.5}] file {Figures_data/EDC_only_256QAM_SD_BW_MI.txt};
\addlegendentry{SD-SW};
\addplot[color=green,thin, mark=o,mark size=1.8,mark options={fill=white,line width=0.5}] file {Figures_data/EDC_only_256QAM_SD_SW_MI.txt};
\addplot[color=red, mark=*,mark options={fill=red,line width=0.5}, mark size=2.2] coordinates {(2300,9.6)}; 

\draw[thick] (13pt,81pt) ellipse (0.06cm and 0.58cm);
\draw[->] (13pt,65pt) -- (18pt,65pt);
\node at (26pt,65pt) {HD};
\draw[thick] (19pt,98pt) ellipse (0.03cm and 0.13cm);
\draw[->] (19pt,102pt) -- (29pt,102pt);
\node at (39pt,102pt) {SD};
\end{axis}
\end{tikzpicture}}
\end{tabular}
\caption{AIRs vs. distance for EDC.}
\label{fig:EDC_AIR}
\end{figure*}
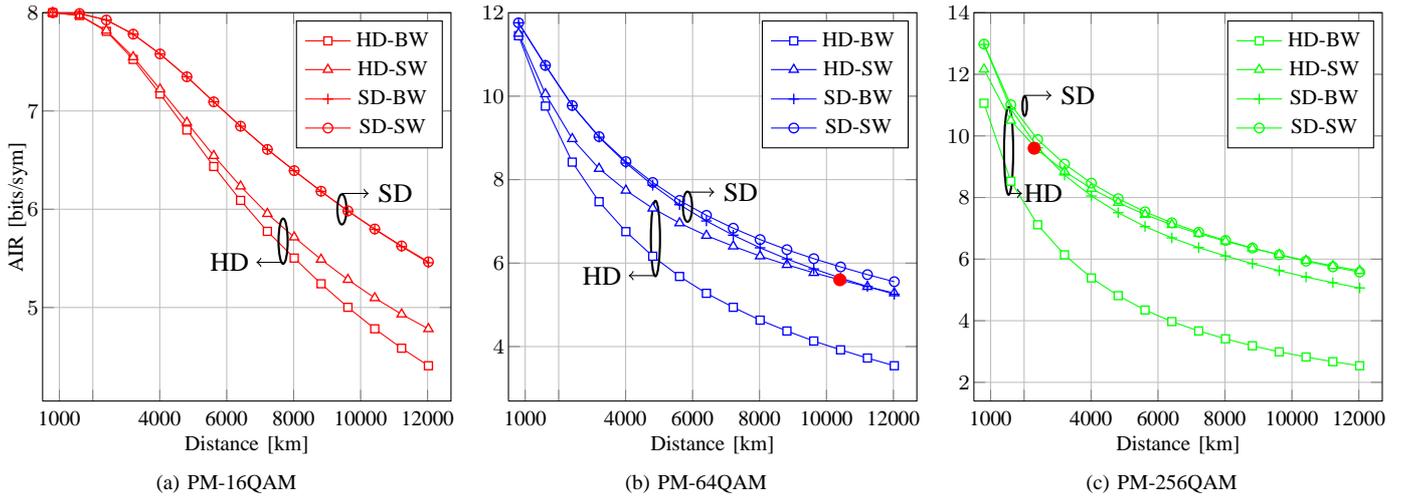

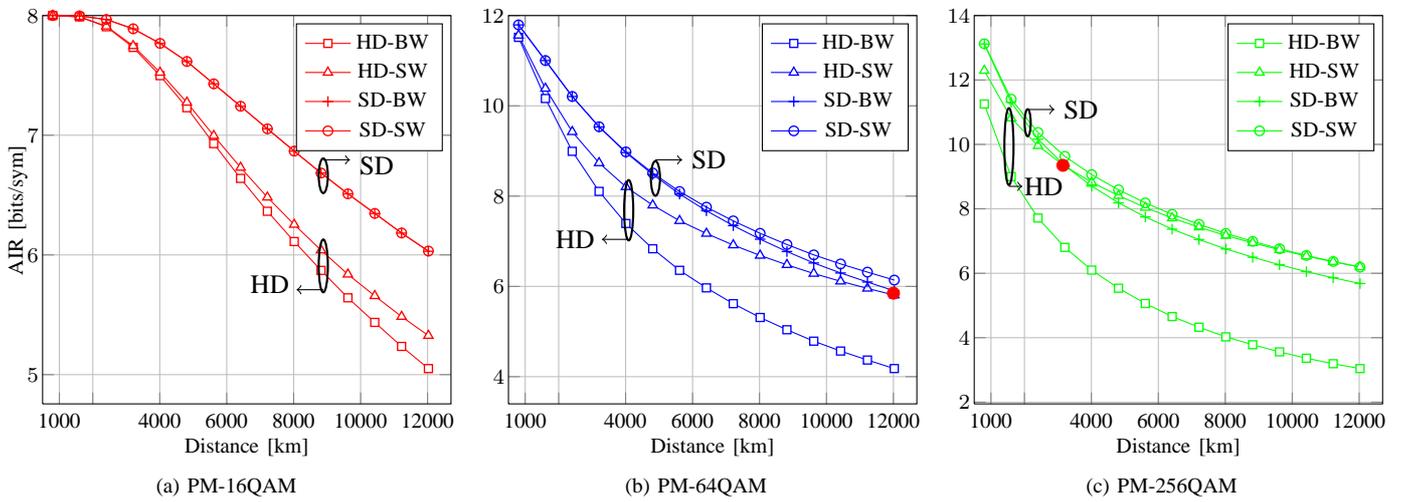
\begin{figure*}[!htbp]
\hspace{-.45cm}
\setlength{\tabcolsep}{0.001cm}
\begin{tabular}{ccc}
\subfloat[PM-16QAM]{
\label{subfig:SC_16QAM}
\begin{tikzpicture}
\begin{axis}[
	legend style={at={(0.98,0.98)},font=\footnotesize,
	legend style={row sep=0.05pt}},
	ticklabel style = {font=\footnotesize},
	xtick={1000, 2000, 4000, 6000, 8000, 10000, 12000},
xticklabels={1000,, 4000, 6000, 8000, 10000,12000},
	width=0.785\columnwidth,
	height=0.76\columnwidth,
	grid=both,
	xmin=500,xmax=12700,
	ymax=8,  
	xlabel={\footnotesize Distance [km]},
	xlabel shift= -3 pt,
	ylabel={\footnotesize AIR [bits/sym]},
	ylabel shift = -5 pt,
	yticklabel shift = -2 pt]
\addlegendentry{HD-BW};
\addplot[color=red,thin, mark=square*,mark size=1.5,mark options={fill=white,line width=0.5}] file {Figures_data/SingleChannelDBP_16QAM_HD_BW_MI.txt};
\addlegendentry{HD-SW};
\addplot[color=red,thin, mark=triangle*,mark size=2,mark options={fill=white,line width=0.5}] file {Figures_data/SingleChannelDBP_16QAM_HD_SW_MI.txt};
\addlegendentry{SD-BW};
\addplot[color=red,thin, mark=+,mark size=2,mark options={fill=white,line width=0.5}] file {Figures_data/SingleChannelDBP_16QAM_SD_BW_MI.txt};
\addlegendentry{SD-SW};
\addplot[color=red,thin, mark=o, mark size=1.8,mark options={fill=white,line width=0.5}] file {Figures_data/SingleChannelDBP_16QAM_SD_SW_MI.txt};
\end{axis}
\draw[thick] (105pt,86pt) ellipse (0.06cm and 0.23cm);
\draw[->] (105pt,92pt) -- (115pt,92pt);
\node at (125pt,90pt) {SD};
\draw[thick] (105pt,52pt) ellipse (0.06cm and 0.35cm);
\draw[->] (105pt,43pt) -- (95pt,43pt);
\node at (85pt,45pt) {HD};
\end{tikzpicture}} &

\subfloat[PM-64QAM]{
\label{subfig:SC_64QAM}
\begin{tikzpicture}
\begin{axis}[
	legend style={at={(0.98,0.98)},font=\footnotesize,
	legend style={row sep=0.05pt}},
	ticklabel style = {font=\footnotesize},
	xtick={1000, 2000, 4000, 6000, 8000, 10000, 12000},
xticklabels={1000,, 4000, 6000, 8000, 10000,12000},
	width=0.785\columnwidth,
	height=0.76\columnwidth,
	grid=both,
	xmin=500,xmax=12700,
	ymax=12,  
	xlabel={\footnotesize Distance [km]},
	xlabel shift= -3 pt,
	ylabel shift = -5 pt,
	yticklabel shift = -2 pt]
\addlegendentry{HD-BW};
\addplot[color=blue,thin, mark=square*,mark size=1.5,mark options={fill=white,line width=0.5}] file {Figures_data/SingleChannelDBP_64QAM_HD_BW_MI.txt};
\addlegendentry{HD-SW};
\addplot[color=blue,thin, mark=triangle*,mark size=2,mark options={fill=white,line width=0.5}] file {Figures_data/SingleChannelDBP_64QAM_HD_SW_MI.txt};
\addlegendentry{SD-BW};
\addplot[color=blue,thin, mark=+,mark size=2,mark options={fill=white,line width=0.5}] file {Figures_data/SingleChannelDBP_64QAM_SD_BW_MI.txt};
\addlegendentry{SD-SW};
\addplot[color=blue,thin, mark=o, mark size=1.8,mark options={fill=white,line width=0.5}] file {Figures_data/SingleChannelDBP_64QAM_SD_SW_MI.txt};
\addplot[color=red, mark=*,mark options={fill=red,line width=0.5}, mark size=2.2] coordinates {(12000,5.85)}; 
\end{axis}
\draw[thick] (45pt,73pt) ellipse (0.06cm and 0.4cm);
\draw[->] (45pt,62pt) -- (35pt,62pt);
\node at (25pt,62pt) {HD};
\draw[thick] (55pt,85pt) ellipse (0.06cm and 0.23cm);
\draw[->] (55pt,92pt) -- (65pt,92pt);
\node at (75pt,92pt) {SD};
\end{tikzpicture}} &
\subfloat[PM-256QAM]{
\label{subfig:SC_256QAM}
\begin{tikzpicture}
\begin{axis}[
	legend style={at={(0.98,0.98)},font=\footnotesize,
	legend style={row sep=0.05pt}},
	ticklabel style = {font=\footnotesize},
	xtick={1000, 2000, 4000, 6000, 8000, 10000, 12000},
xticklabels={1000, ,4000, 6000, 8000, 10000,12000},
	width=0.785\columnwidth,
	height=0.76\columnwidth,
	grid=both,
	xmin=500,xmax=12700,
	ymax=14,  
	xlabel={\footnotesize Distance [km]},
	xlabel shift= -3 pt,
	ylabel shift = -5 pt,
	yticklabel shift = -2 pt]
\addlegendentry{HD-BW};
\addplot[color=green,thin, mark=square*,mark size=1.5,mark options={fill=white,line width=0.5}] file {Figures_data/SingleChannelDBP_256QAM_HD_BW_MI.txt};
\addlegendentry{HD-SW};
\addplot[color=green,thin, mark=triangle*,mark size=2,mark options={fill=white,line width=0.5}] file {Figures_data/SingleChannelDBP_256QAM_HD_SW_MI.txt};
\addlegendentry{SD-BW};
\addplot[color=green,thin, mark=+,mark size=2,mark options={fill=white,line width=0.5}] file {Figures_data/SingleChannelDBP_256QAM_SD_BW_MI.txt};
\addlegendentry{SD-SW};
\addplot[color=green,thin, mark=o, mark size=1.8,mark options={fill=white,line width=0.5}] file {Figures_data/SingleChannelDBP_256QAM_SD_SW_MI.txt};
\addplot[color=red, mark=*,mark options={fill=red,line width=0.5}, mark size=2.2] coordinates {(3150,9.35)};
\end{axis}
\draw[thick] (13pt,97pt) ellipse (0.06cm and 0.51cm);
\draw[->] (13pt,82pt) -- (18pt,82pt);
\node at (26pt,82pt) {HD};
\draw[thick] (20pt,106pt) ellipse (0.04cm and 0.18cm);
\draw[->] (20pt,111pt) -- (30pt,111pt);
\node at (40pt,111pt) {SD};
\end{tikzpicture}}
\end{tabular}
\caption{AIRs vs. distance for single-channel DBP.}
\label{fig:SingleChDBP}
\end{figure*}

\begin{figure*}[!htbp]
\hspace{-.45cm}
\setlength{\tabcolsep}{0.001cm}
\begin{tabular}{ccc}
\subfloat[PM-16QAM]{
\begin{tikzpicture}
\begin{axis}[
	legend style={at={(0.4,0.4)},font=\footnotesize,
	legend style={row sep=0.05pt}},
	ticklabel style = {font=\footnotesize},
	xtick={1000, 2000, 4000, 6000, 8000, 10000, 12000},
xticklabels={1000,, 4000, 6000, 8000, 10000,12000},
ytick={7,7.5,8},
yticklabels={7,,8},
	width=0.785\columnwidth,
	height=0.76\columnwidth,
	grid=both,
	xmin=500,xmax=12700,
	ymax=8,  
	xlabel={\footnotesize Distance [km]},
	xlabel shift= -3 pt,
	ylabel={\footnotesize AIR [bits/sym]},
	ylabel shift = -5 pt,
	yticklabel shift = -2 pt]
\addlegendentry{HD-BW};
\addplot[color=red,thin, mark=square*,mark size=1.5,mark options={fill=white,line width=0.5}] file {Figures_data/FullFieldDBP_16QAM_HD_BW_MI.txt};
\addlegendentry{HD-SW};
\addplot[color=red,thin, mark=triangle*,mark size=2,mark options={fill=white,line width=0.5}] file {Figures_data/FullFieldDBP_16QAM_HD_SW_MI.txt};
\addlegendentry{SD-BW};
\addplot[color=red,thin, mark=+,mark size=2,mark options={fill=white,line width=0.5}] file {Figures_data/FullFieldDBP_16QAM_SD_BW_MI.txt};
\addlegendentry{SD-SW};
\addplot[color=red,thin, mark=o, mark size=1.8,mark options={fill=white,line width=0.5}] file {Figures_data/FullFieldDBP_16QAM_SD_SW_MI.txt};
\end{axis}

\draw[thick] (125pt,63pt) ellipse (0.06cm and 0.4cm);
\draw[->] (125pt,52pt) -- (115pt,52pt);
\node at (105pt,52pt) {HD};
\draw[thick] (120pt,113pt) ellipse (0.06cm and 0.23cm);
\draw[->] (120pt,120pt) -- (130pt,120pt);
\node at (140pt,120pt) {SD};
\end{tikzpicture}\label{subfig:FF_16QAM}} &

\subfloat[PM-64QAM]{
\begin{tikzpicture}
\begin{axis}[
	legend style={at={(0.98,0.98)},font=\footnotesize,
	legend style={row sep=0.05pt}},
	ticklabel style = {font=\footnotesize},
	xtick={1000, 2000, 4000, 6000, 8000, 10000, 12000},
xticklabels={1000,, 4000, 6000, 8000, 10000,12000},
	width=0.785\columnwidth,
	height=0.76\columnwidth,
	grid=both,
	xmin=500,xmax=12700,
	ymax=12,  
	xlabel={\footnotesize Distance [km]},
	xlabel shift= -3 pt,
	ylabel shift = -5 pt,
	yticklabel shift = -2 pt]
\addlegendentry{HD-BW};
\addplot[color=blue,thin, mark=square*,mark size=1.5,mark options={fill=white,line width=0.5}] file {Figures_data/FullFieldDBP_64QAM_HD_BW_MI.txt};
\addlegendentry{HD-SW};
\addplot[color=blue,thin, mark=triangle*,mark size=2,mark options={fill=white,line width=0.5}] file {Figures_data/FullFieldDBP_64QAM_HD_SW_MI.txt};
\addlegendentry{SD-BW};
\addplot[color=blue,thin, mark=+,mark size=2,mark options={fill=white,line width=0.5}] file {Figures_data/FullFieldDBP_64QAM_SD_BW_MI.txt};
\addlegendentry{SD-SW};
\addplot[color=blue,thin, mark=o, mark size=1.8,mark options={fill=white,line width=0.5}] file {Figures_data/FullFieldDBP_64QAM_SD_SW_MI.txt};
\end{axis}
\draw[thick] (105pt,55pt) ellipse (0.06cm and 0.5cm);
\draw[->] (105pt,41pt) -- (95pt,41pt);
\node at (85pt,41pt) {HD};
\draw[thick] (105pt,83pt) ellipse (0.06cm and 0.23cm);
\draw[->] (105pt,90pt) -- (115pt,90pt);
\node at (125pt,90pt) {SD};
\end{tikzpicture}\label{subfig:FF_64QAM}} &

\subfloat[PM-256QAM]{
\begin{tikzpicture}
\begin{axis}[
	legend style={at={(0.98,0.98)},font=\footnotesize,
	legend style={row sep=0.05pt}},
	ticklabel style = {font=\footnotesize},
	xtick={1000, 2000, 4000, 6000, 8000, 10000, 12000},
xticklabels={1000,, 4000, 6000, 8000, 10000,12000},
	width=0.785\columnwidth,
	height=0.76\columnwidth,
	grid=both,
	xmin=500,xmax=12700,
	ymax=16,  
	xlabel={\footnotesize Distance [km]},
	xlabel shift= -3 pt,
	ylabel shift = -5 pt,
	yticklabel shift = -2 pt]
\addlegendentry{HD-BW};
\addplot[color=green,thin, mark=square*,mark size=1.5,mark options={fill=white,line width=0.5}] file {Figures_data/FullFieldDBP_256QAM_HD_BW_MI.txt};
\addlegendentry{HD-SW};
\addplot[color=green,thin, mark=triangle*,mark size=2,mark options={fill=white,line width=0.5}] file {Figures_data/FullFieldDBP_256QAM_HD_SW_MI.txt};
\addlegendentry{SD-BW};
\addplot[color=green,thin, mark=+,mark size=2,mark options={fill=white,line width=0.5}] file {Figures_data/FullFieldDBP_256QAM_SD_BW_MI.txt};
\addlegendentry{SD-SW};
\addplot[color=green,thin, mark=o, mark size=1.8,mark options={fill=white,line width=0.5}] file {Figures_data/FullFieldDBP_256QAM_SD_SW_MI.txt};
\end{axis}
\draw[thick] (55pt,77pt) ellipse (0.06cm and 0.50cm);
\draw[->] (55pt,63pt) -- (45pt,63pt);
\node at (35pt,63pt) {HD};
\draw[thick] (45pt,107pt) ellipse (0.06cm and 0.23cm);
\draw[->] (45pt,114pt) -- (55pt,114pt);
\node at (65pt,114pt) {SD};
\end{tikzpicture}\label{subfig:FF_256QAM}}
\end{tabular}
\caption{AIRs vs. distance with full-field DBP.}
\label{fig:FullFieldDBP}
\end{figure*}
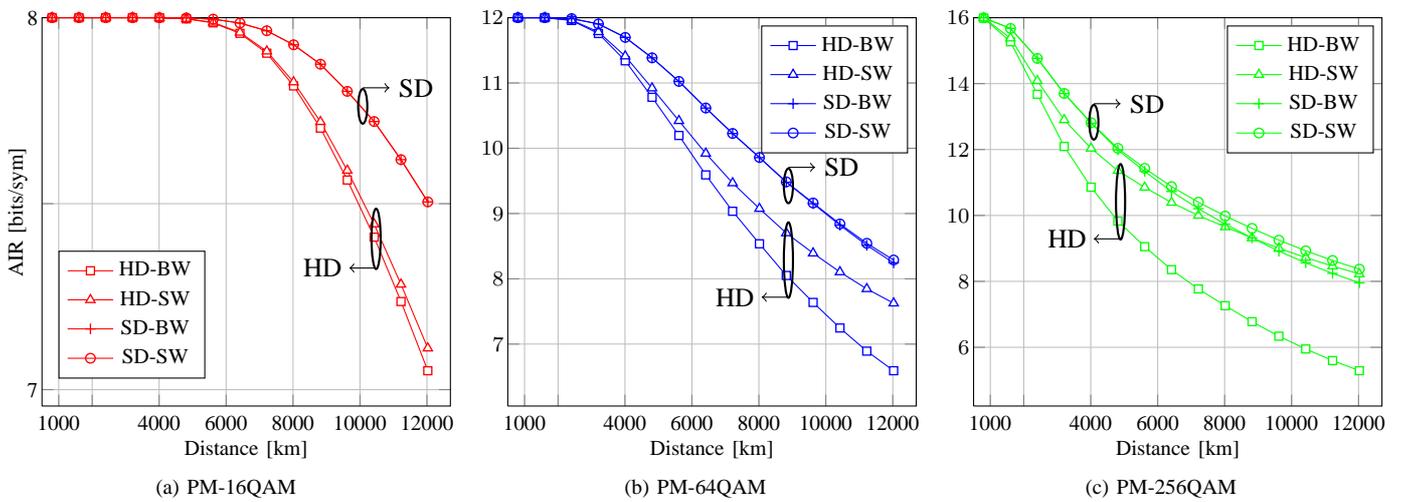

In Figs.~\ref{fig:EDC_AIR}--\ref{fig:FullFieldDBP}, three sets of results on AIRs for the optical fiber channel are shown: EDC, single-channel DBP, and full-field DBP, respectively. Each set shows the AIR vs. transmission distance for PM-16QAM, PM-64QAM, and PM-256QAM with the four CM decoder structures discussed in Section~\ref{sec2}. For each distance, equalization scheme, and CM decoder investigated, the transmitted power was optimized, resulting in different optimal powers. The investigated link distances span the typical distances of mid-range to long-haul terrestrial links (typically 1000--3000 km), long-haul submarine (3000--5000 km), and transoceanic links (6000--12000 km). 
\begin{table}[!t]
\caption{System parameters} 
\label{tab:syspar}
\centering
\begin{tabular}{|c|c|}
\hline
{\bf Parameter Name} & {\bf Value} \\
\hline \hline
\multicolumn{2}{|c|}{\emph{Transmitter Parameters}} \\ 
\hline
WDM Channels & 5 \\
\hline
Symbol Rate & 32 Gbaud \\
\hline
RRC Roll-Off & 0.01\\
\hline
Channel Frequency Spacing & 33 GHz \\
\hline\hline
\multicolumn{2}{|c|}{\emph{Fiber Channel Parameters}}  \\
\hline
Attenuation ($\alpha$) & 0.2 dB/km \\
\hline
Dispersion Parameter ($D$) & 17 ps/nm/km \\
\hline
Nonlinearity Parameter ($\gamma$) & 1.2 1/(W$\cdot$km) \\
\hline
Fiber Span Length & 80 km \\
\hline
EDFA Gain &  16 dB \\
\hline
EDFA Noise Figure & 4.5 dB \\
\hline\hline
\multicolumn{2}{|c|}{\emph{Numerical Parameters}}  \\
\hline
SSF Spatial Step Size & 100 m \\
\hline
Simulation Bandwidth & 512 GHz \\
\hline
\end{tabular}
\end{table}

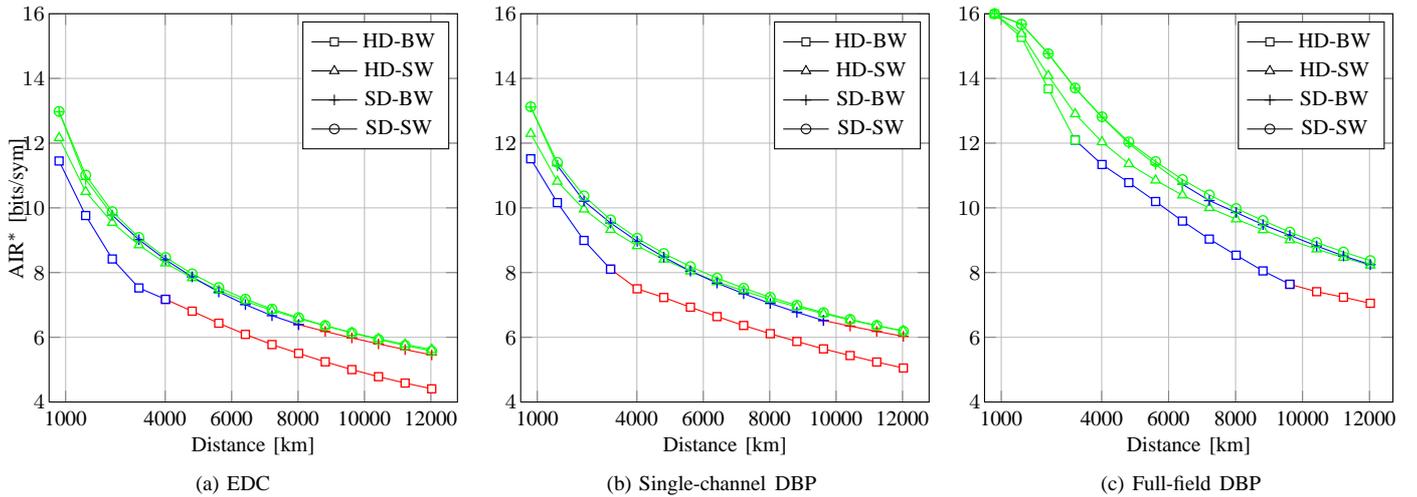
\begin{figure*}[!htbp]
\hspace{-.45cm}
\setlength{\tabcolsep}{0.0001cm}
\begin{tabular}{ccc}
\subfloat[EDC]{
\begin{tikzpicture}
\begin{axis}[legend style={at={(0.98,0.98)},font=\footnotesize,
	legend style={row sep=0.05pt}},
	ticklabel style = {font=\footnotesize},
	xtick={1000, 4000, 6000, 8000, 10000, 12000},
    xticklabels={1000, 4000, 6000, 8000, 10000,12000},
	width=0.785\columnwidth,
	height=0.76\columnwidth,
	grid=both,
	xmin=500,xmax=12800,
	ymin=4,
	ymax=16,  
	xlabel={\footnotesize Distance [km]},
	xlabel shift= -3 pt,
	ylabel={\footnotesize AIR$^*$ [bits/sym]},
	ylabel shift = -7pt,
	yticklabel shift = -2 pt] 
	
	\addlegendentry{HD-BW};
	\addlegendimage{mark=square*,mark size=1.5,mark options={fill=white},black};
    \addlegendentry{HD-SW};
    \addlegendimage{mark=triangle*,mark size=2,mark options={fill=white},black};
    \addlegendentry{SD-BW};
    \addlegendimage{mark=+,mark size=2,mark options={fill=white},black};
    \addlegendentry{SD-SW};
    \addlegendimage{mark=o,mark size=1.8,mark options={fill=white},black};
    
\addplot[color=red,thin, mark=square*,mark size=1.5,mark options={fill=white,line width=0.5}] file {Figures_data/EDC_only_opt_HD_BW_16QAM.txt};
\addplot[color=blue,thin,mark=square*,mark size=1.5,mark options={fill=white,line width=0.5}] file {Figures_data/EDC_only_opt_HD_BW_64QAM.txt};
\addplot[color=green,thin,mark=triangle*,mark size=2,mark options={fill=white,line width=0.5}] file {Figures_data/EDC_only_opt_HD_SW_256QAM.txt};
\addplot[color=red,thin,mark=+,mark size=2,mark options={fill=white,line width=0.5}] file {Figures_data/EDC_only_opt_SD_BW_16QAM.txt};
\addplot[color=blue, mark=+,mark size=2,mark options={fill=white,line width=0.5}] file {Figures_data/EDC_only_opt_SD_BW_64QAM.txt};
\addplot[color=green,thin, mark=+,mark size=2,mark options={fill=white,line width=0.5}] file {Figures_data/EDC_only_opt_SD_BW_256QAM.txt};
\addplot[color=green,thin, mark=o,mark size=1.8,mark options={fill=white,line width=0.5}] file {Figures_data/EDC_only_opt_SD_SW_256QAM.txt};
\end{axis}
\end{tikzpicture}
\label{subfig:EDC_only}} 
&
\subfloat[Single-channel DBP]{
\begin{tikzpicture}
\begin{axis}[
	legend style={at={(0.98,0.98)},font=\footnotesize,
	legend style={row sep=0.05pt}},
	ticklabel style = {font=\footnotesize},
	xtick={1000, 4000, 6000, 8000, 10000, 12000},
xticklabels={1000, 4000, 6000, 8000, 10000,12000},
	width=0.785\columnwidth,
	height=0.76\columnwidth,
	grid=both,
	xmin=500,xmax=12800,
	ymin=4,
	ymax=16,  
	xlabel={\footnotesize Distance [km]},
	xlabel shift= -3 pt,
	ylabel shift = -5 pt,
	yticklabel shift = -2 pt]
	
    \addlegendentry{HD-BW};
	\addlegendimage{mark=square*,mark size=1.5,mark options={fill=white},black};
    \addlegendentry{HD-SW};
    \addlegendimage{mark=triangle*,mark size=2,mark options={fill=white},black};
    \addlegendentry{SD-BW};
    \addlegendimage{mark=+,mark size=2,mark options={fill=white},black};
    \addlegendentry{SD-SW};
    \addlegendimage{mark=o,mark size=1.8,mark options={fill=white},black};	
	
    \addplot[color=red,thin, mark=square*,mark size=1.5,mark options={fill=white,line width=0.5}] file {Figures_data/SingleChannelDBP_opt_HD_BW_16QAM.txt};
\addplot[color=blue,thin,mark=square*,mark size=1.5,mark options={fill=white,line width=0.5}] file {Figures_data/SingleChannelDBP_opt_HD_BW_64QAM.txt};
\addplot[color=green,thin,mark=triangle*,mark size=2,mark options={fill=white,line width=0.5}] file {Figures_data/SingleChannelDBP_opt_HD_SW_256QAM.txt};
\addplot[color=red,thin,mark=+,mark size=2,mark options={fill=white,line width=0.5}] file {Figures_data/SingleChannelDBP_opt_SD_BW_16QAM.txt};
\addplot[color=blue, mark=+,mark size=2,mark options={fill=white,line width=0.5}] file {Figures_data/SingleChannelDBP_opt_SD_BW_64QAM.txt};
\addplot[color=green,thin, mark=+,mark size=2,mark options={fill=white,line width=0.5}] file {Figures_data/SingleChannelDBP_opt_SD_BW_256QAM.txt};
\addplot[color=green,thin, mark=o,mark size=1.8,mark options={fill=white,line width=0.5}] file {Figures_data/SingleChannelDBP_opt_SD_SW_256QAM.txt};	
\end{axis}
\end{tikzpicture}
\label{subfig:SingleChDBP}}
&
\subfloat[Full-field DBP]{
\hspace{-.23cm}
\begin{tikzpicture}
\begin{axis}[
	legend style={at={(0.98,0.98)},font=\footnotesize,
	legend style={row sep=0.05pt}},
	ticklabel style = {font=\footnotesize},
	xtick={1000, 4000, 6000, 8000, 10000, 12000},
xticklabels={1000, 4000, 6000, 8000, 10000,12000},
	width=0.785\columnwidth,
	height=0.76\columnwidth,
	grid=both,
	xmin=500,xmax=12700,
	ymin=4,
	ymax=16,  
	xlabel={\footnotesize Distance [km]},
	xlabel shift= -3 pt,
	ylabel shift = -5 pt,
	yticklabel shift = -2 pt]
	
	\addlegendentry{HD-BW};
	\addlegendimage{mark=square*,mark size=1.5,mark options={fill=white},black};
    \addlegendentry{HD-SW};
    \addlegendimage{mark=triangle*,mark size=2,mark options={fill=white},black};
    \addlegendentry{SD-BW};
    \addlegendimage{mark=+,mark size=2,mark options={fill=white},black};
    \addlegendentry{SD-SW};
    \addlegendimage{mark=o,mark size=1.8,mark options={fill=white},black};

\addplot[color=red,thin, mark=square*,mark size=1.5,mark options={fill=white,line width=0.5}] file {Figures_data/FullFieldDBP_opt_HD_BW_16QAM.txt};
\addplot[color=blue,thin,mark=square*,mark size=1.5,mark options={fill=white,line width=0.5}] file {Figures_data/FullFieldDBP_opt_HD_BW_64QAM.txt};
\addplot[color=green,thin,mark=square*,mark size=1.5,mark options={fill=white,line width=0.5}] file {Figures_data/FullFieldDBP_opt_HD_BW_256QAM.txt};
\addplot[color=green,thin,mark=triangle*,mark size=2,mark options={fill=white,line width=0.5}] file {Figures_data/FullFieldDBP_opt_HD_SW_256QAM.txt};
\addplot[color=blue, mark=+,mark size=2,mark options={fill=white,line width=0.5}] file {Figures_data/FullFieldDBP_opt_SD_BW_64QAM.txt};
\addplot[color=green,thin, mark=+,mark size=2,mark options={fill=white,line width=0.5}] file {Figures_data/FullFieldDBP_opt_SD_BW_256QAM.txt};
\addplot[color=green,thin, mark=o,mark size=1.8,mark options={fill=white,line width=0.5}] file {Figures_data/FullFieldDBP_opt_SD_SW_256QAM.txt};
\end{axis}
\end{tikzpicture}
\label{subfig:FullFieldDBP}}
\end{tabular}
\caption{AIRs vs. distance for the optimal PM-$M$QAM format, indicated by red ($M=16$), blue ($M=64$) and green ($M=256$).}\label{fig:OptFormats}
\end{figure*}  

In the EDC case for PM-16QAM (Fig.~\ref{subfig:EDC_16QAM}), SD decoders significantly outperform the HD ones, particularly for long distances. SD-BW decoders incur small penalties compared to the SD-SW implementation at all distances of interest. This can be explained by observing Fig.~\subref*{subfig:16QAM}, where the performance of PM-16QAM differs for SD-SW and SD-BW decoders only for very small SNR values ($\leq$2 dB). As shown in Fig.~\ref{subfig:EDC_64QAM}, for the PM-64QAM format, SD decoders show a significant advantage over their HD counterparts (see \cite{Fehenberger2015} for SD-SW vs. HD-BW) and again SD-BW decoders have identical performance as the SD-SW ones at short distances. However, as the distance is increased, the AIRs of the HD-SW schemes match the SD-BW ones (see filled red circles in Fig.~\subref*{subfig:EDC_64QAM} and \subref*{subfig:EDC_256QAM}), significantly outperforming the HD-BW rates. This trend is even more prominent for PM-256QAM (Fig.~\subref*{subfig:EDC_256QAM}). For this format, a crossing between the SD-BW and HD-SW AIRs can be observed at around 2300 km distance (filled red circles). More importantly, in the long distance regime, the HD-SW scheme matches the performance of the SD-SW one, with no significant penalty observed. Also, it can be noted that the HD-BW scheme shows a significant penalty ($>$3 bits/sym for long distances) compared to all other implementations. 
 
In the case where single-channel DBP is applied (Fig.~\ref{fig:SingleChDBP}), rather small AIR gains can be noticed in general, as compared to the EDC case (Fig.~\ref{fig:EDC_AIR}). This can be attributed to the fact that the compensation of the nonlinearity generated by only one channel out of the five transmitted gives only a marginal improvement of the optimum SNR at each transmission distance. However, some differences in the performance can be noticed for higher order formats and long distances. Specifically, the distance at which the HD-SW transceiver matches the performance of the SD-BW ones for PM-64QAM is increased from 10000 km to 12000 km (filled red circles in Fig.~\subref*{subfig:EDC_64QAM} and Fig.~\subref*{subfig:SC_64QAM}) and for PM-256QAM the crossing point between HD-SW and SD-BW is moved from 2300 km to 3000 km (filled red circles in Fig.~\subref*{subfig:EDC_256QAM} and Fig.~\subref*{subfig:SC_256QAM}).

Finally, when full compensation of signal--signal nonlinear distortion is performed via full-field DBP (Fig.~\ref{fig:FullFieldDBP}), a remarkable increase in the AIRs compared to the other equalization schemes can be observed for all decoding strategies and all modulation formats. Fig.~\subref*{subfig:FF_16QAM} shows that, for PM-16QAM, the full nominal SE (8 bits/sym) can be achieved up to a distance of approximately 6000 km and by only using an HD-BW decoder (squares). This rate drops by only 0.5 bits/sym at 12000 km if SD decoders are used, and by an additional 0.5 bits/sym (to 7 bits/sym) when HD decoders are adopted. Fig.~\subref*{subfig:FF_16QAM} also shows that when PM-16QAM and full-field DBP are used in conjunction, switching from a binary to a nonbinary scheme does not result in any significant AIR increase, as long as the FEC decoding strategy (HD or SD) is maintained. Higher rates can be achieved using PM-64QAM (Fig.~\subref*{subfig:FF_64QAM}) and PM-256QAM (Fig.~\subref*{subfig:FF_256QAM}) in conjunction with SD decoders. Again, binary and nonbinary SD schemes perform identically. For these higher order modulation formats, HD-BW decoders incur significant penalties compared to SD decoders. For PM-64QAM, this penalty becomes larger than 0.5 bits/sym for distances larger than 4000 km whereas for PM-256QAM, they become larger than 0.5 bits/sym already for distances larger than 1500 km. At long distances, the penalty increases to up to 1.6 bits/sym for PM-64QAM and 2.5 bits/sym for PM-256QAM. An improvement can be obtained by using HD-SW decoders, particularly in the long-distance regime. For PM-64QAM, the AIR gap from SD decoders is reduced to 0.5 bits/sym at 12000 km. For PM-256QAM, HD-SW decoders in general largely outperform HD-BW decoders and show performances similar to SD decoders beyond distances of 7000 km, outperforming also SD-BW decoders beyond 8000 km. 

In order to highlight the performance of each decoding structure vs. the transmission distance $L$, in Fig.~\ref{fig:OptFormats} we show the modulation format optimized AIRs, defined as 

\begin{equation}
\text{AIR}^*(L)=\max_{M\in\{16,64,256\}}\text{AIR}(L,M)
\end{equation} 
for EDC, single-channel DBP, and full-field DBP.

We observe that the set of curves shown for each equalization scheme appears as a shifted version (across the distance axis) of the other ones. This behavior is another confirmation of the fact that dispersion-unmanaged and EDFA-amplified optical fiber systems can be described by an equivalent AWGN channel and their performance is strongly correlated to the effective SNR at the MF output. Since this SNR includes nonlinear effects as an equivalent noise source, it is improved by nonlinear compensation schemes.
In the EDC case (Fig.~\subref*{subfig:EDC_only}), except for short distances ($\leq$1000 km), HD-SW decoders have comparable performance to SD-BW and SD-SW schemes. The optimal format for both SW strategies (SD and HD) is PM-256QAM (green) at all distances, whereas for the BW schemes, PM-256QAM performs worse both for short and middle distances, where PM-64QAM (blue) is preferable, as well as in the long/ultra-long haul region, where PM-16QAM (red) is optimal. Very similar behavior is observed for single-channel DBP in Fig.~\subref*{subfig:SingleChDBP}, where the optimality of PM-64QAM for BW receivers is extended to longer distances with respect to their EDC counterparts.

Finally, for full-field DBP (Fig.~\subref*{subfig:FullFieldDBP}), rates of up to 12 bits/sym can be targeted up to 5000 km, and for all decoding strategies, the optimal modulation format is PM-256QAM up to 4000 km. Also, in the ultra-long haul regime, rates above 8 bits/sym can be achieved by using PM-64QAM in conjunction with SD-BW systems without significant loss in performance compared to SD-SW or HD-SW with PM-256QAM.  
Overall, Fig.~\ref{fig:OptFormats} also shows that HD-BW decoders perform significantly worse than all other schemes, confirming the results in \cite{Fehenberger2015}. Nevertheless, they can be considered as a valid low-complexity alternative for short distances or when high SNRs are available at the receiver.    

\section{Conclusions}\label{sec5}
The MI is a useful measure of  performance of a coded system and represents an upper bound on the AIRs when a given modulation format is used and optimum ML decoding is performed at the receiver. Conversely, the AIRs of pragmatic transceiver schemes are  dictated by the specific implementation of the CM decoder. In this work, we presented a detailed numerical study of the AIR performance for high-SE long-haul optical communication systems when these pragmatic decoders and equalization schemes, such as EDC and DBP are employed. 

The results in this paper lead to  interesting conclusions on the performance of coded optical fiber communication systems. For example, when the equalizer enables high SNR values (through the use of full-field DBP), an SD decoder is not the only alternative to achieve high rates at long distances. On the contrary, HD nonbinary FEC schemes can, in principle, achieve almost the same rates across all distances of interest. 
For SNR values in the low to medium range (EDC or single-channel DBP), SD decoders outperform HD ones up to medium SE formats (PM-64QAM). However, for high-SE formats (PM-256QAM), the HD-SW CM decoder can outperform the SD-BW decoder. In the SD case, BW decoders do not incur significant penalties as compared to their SW counterparts, suggesting that there is no need to employ nonbinary FEC schemes. Finally, HD-BW transceivers are never desirable for high-SE systems. Nevertheless, they can represent the implementation of choice for either short-distance systems or ultra long-haul low-SE systems whenever high order modulation formats cannot be used.   
\balance

\ifCLASSOPTIONcaptionsoff
  \newpage
\fi



\bibliographystyle{IEEEtran}
\bibliography{biblio}
%

%







\end{document}